\documentclass[11pt]{article}
\usepackage{enumerate}
\usepackage{enumitem}
\usepackage{fixltx2e}
\usepackage{graphicx}
\usepackage{longtable}
\usepackage{float}
\usepackage{setspace}
\usepackage{wrapfig}
\usepackage{soul}
\usepackage{amsmath}
\usepackage{mathtools}
\usepackage{mathrsfs}
\usepackage{textcomp}
\usepackage{marvosym}
\usepackage{wasysym}
\usepackage{latexsym}
\usepackage{amssymb}
\usepackage{hyperref}
\usepackage{graphicx}
\usepackage{subfig}
\usepackage{longtable}
\usepackage{float}
\usepackage{array}
\usepackage[ruled]{algorithm}
\usepackage[noend]{algpseudocode}
\usepackage[table]{xcolor}
\usepackage{arydshln}
\usepackage{cancel}
\usepackage{url}
\usepackage{comment}
\usepackage{setspace}
\usepackage[margin=1in]{geometry}
\usepackage[affil-it]{authblk}
\usepackage{titlesec}
\usepackage{color}
\RequirePackage{latexsym}
\RequirePackage{amsmath}
\RequirePackage{amssymb}
\RequirePackage{bm}             %
\RequirePackage{soul}
\usepackage{amsmath}
\usepackage{graphicx,psfrag,epsf}
\usepackage{enumerate}
\usepackage{natbib}
\usepackage{url} 

\newcommand{\blind}{1}

\newtheorem{algor}{Scheme}
\newcommand{\ema}{\text{EM}}
\newcommand{\dem}{\text{DEM}}

\DeclareMathOperator{\ab}{\mathbf{a}}
\DeclareMathOperator{\bb}{\mathbf{b}}

\DeclareMathOperator{\eb}{\mathbf{e}}

\DeclareMathOperator{\yb}{\mathbf{y}}

\DeclareMathOperator{\Lcal}{\mathcal{L}}
\DeclareMathOperator{\Mcal}{\mathcal{M}}

\DeclareMathOperator{\Scal}{\mathcal{S}}

\DeclareMathOperator{\EE}{\mathbb{E}} 
\DeclareMathOperator{\RR}{\mathbb{R}} 

\DeclareMathOperator{\zero}{\mathbf{0}} 

\DeclareMathOperator*{\argmax}{\mathop{\mathrm{argmax}}}

\DeclareMathOperator{\diag}{diag}

\DeclareMathOperator{\betab}{{\bm{\beta}}}

\ifx\BlackBox\undefined
\newcommand{\BlackBox}{\rule{1.5ex}{1.5ex}}  
\fi

\newtheorem{theorem}{Theorem}[section]

\ifx\proof\undefined
\newenvironment{proof}{\par\noindent{\bf Proof\ }}{\hfill\BlackBox\\[2mm]}
\fi

\ifx\axiom\undefined
\newtheorem{axiom}[theorem]{Axiom}
\fi

\begin{document}

\if1\blind
{
\title{An Asynchronous Distributed Expectation Maximization Algorithm For Massive Data: The DEM Algorithm}
\author{Sanvesh Srivastava \thanks{Department of Statistics and Actuarial Science, The University of Iowa, Iowa City, Iowa 52242, \url{sanvesh-srivastava@uiowa.edu}},
  Glen DePalma \thanks{Department of Statistics, Purdue University, West Lafayette, Indiana 47907, \url{glen.depalma@gmail.com}} \&
  Chuanhai Liu \thanks{Department of Statistics, Purdue University, West Lafayette, Indiana 47907, \url{chuanhai@purdue.edu}}}
  \maketitle
} \fi

\if0\blind
{
  \bigskip
  \bigskip
  \bigskip
  \begin{center}
   {\LARGE\bf An Asynchronous Distributed Expectation Maximization Algorithm For Massive Data: The DEM Algorithm}
  \end{center}
  \medskip
} \fi

\bigskip
\begin{abstract}
  The family of Expectation-Maximization (EM) algorithms provides a general approach to fitting flexible models for large and complex data. The expectation (E) step of EM-type algorithms is time consuming in massive data applications because it requires multiple passes through the full data. We address this problem by proposing an asynchronous and distributed generalization of the EM called the Distributed EM (DEM). Using DEM, existing EM-type algorithms are easily extended to massive data settings by exploiting the divide-and-conquer technique and widely available computing power, such as grid computing. The DEM algorithm reserves two groups of computing processes called \emph{workers} and \emph{managers} for  performing the E step and the maximization step (M step), respectively. The samples are randomly partitioned into a large number of disjoint subsets and are stored on the worker processes. The E step of DEM algorithm is performed in parallel on all the workers, and every worker communicates its results to the managers at the end of local E step. The managers perform the M step after they have received results from a $\gamma$-fraction of the workers, where $\gamma$ is a fixed constant in $(0, 1]$. The sequence of parameter estimates generated by the DEM algorithm retains the attractive properties of EM: convergence of the sequence of parameter estimates to a local mode and linear global rate of convergence. Across diverse simulations focused on linear mixed-effects models, the DEM algorithm is significantly faster than competing EM-type algorithms while having a similar accuracy. The DEM algorithm maintains its superior empirical performance on a movie ratings database consisting of 10 million ratings. 
\end{abstract}

\noindent%

\section{Introduction}
\label{sec:intro}

Developing efficient generalizations of the EM algorithm is an active area of research. The monotonic ascent of the EM algorithm can be extremely slow, especially near the optimum. There are many EM extensions that increase the speed of convergence of EM while retaining its stability and simplicity and that reduce to the original EM algorithm under certain assumptions. EM-type algorithms are very slow in massive data applications because the E step requires multiple passes through the whole data for each iteration. Memory limitations further worsen efficiency of the E step. This has motivated a rich literature on online EM algorithms, which modify the E step using stochastic approximation. Our goal is to propose an asynchronous and distributed generalization of the EM called the DEM algorithm. It leads to easy extensions of EM-type algorithms using the divide-and-conquer technique. Distributed computations allow scalability to arbitrarily large data sets, and asynchronous computations minimize communication cost for each iteration. Both features are key in maintaining the efficiency of DEM algorithm in massive data applications while retaining the simplicity and stability of EM-type algorithms. 

The EM algorithm has been extended by generalizing the missing data augmentation schemes or by developing efficient M steps. In most of these extensions, the E step uses data from every sample. Such extensions are inefficient in massive data settings for two main reasons. First, every iteration is time consuming due to a large number of samples. Second, if the data require many machines for storage, then extensive communication among all the machines further increases the time of each iteration; therefore, EM \citep{DemLaiRub77} and the family of EM-type algorithms, such as ECM, ECME, AECM, PXEM, and DECME \citep{MenRub93,LiuRub94,MenDyk97,LiuRubWu98,HeLiu12}, are inefficient in massive data settings simply due to the time consuming E step or possibly due to the communication cost. The same is also true for EM extensions that modify the M step by borrowing ideas from optimization \citep{Lan95,JamJen97,NeaHin98,SalRow03,VarRol08,Yu12}.

Current EM extensions for massive data applications are based on stochastic approximation and fall in the online EM family \citep{Tit84,Lan95}. \citet{CapMou09} generalized online EMs to statistical models that have their complete-data likelihood in the curved exponential family. All online EMs use the data sequentially rather than in a batch and continuously update parameter estimates as data arrive. Online EMs modify the E step of the classical EM to an online E step that computes the conditional expectation of the complete-data log likelihood obtained using a small fraction of the full data. As greater fraction of the full data are processed, the online E step increases in accuracy and yields similar results as the E step of classical EM. The M step of an online EM is same as that of the classical EM. Online EMs retain the simplicity of implementation of the classical EM but fail to retain the monotone ascent of the likelihood for each iteration. Developing online EMs for complex models is an active area of research \citep{Cap11,Leetal11}.

EM extensions based on the divide-and-conquer technique provide an alternative to online EMs in massive data settings. The methods in this class divide the data into smaller $K$ disjoint subsets and perform E steps in parallel on the $K$ subsets. The E steps could be performed in parallel on different nodes in a cluster, threads of a graphical processing unit (GPU), or processors in a computer, which are generically called processes. The results of all the parallel E steps are combined into an objective for maximization in the M step. Typically, all processes communicate the conditional expectations of sufficient statistics to a common process, which combines them before performing the M step. A variety of such algorithms exist for mixture models \citep{Now03,Gu08,ZhoLanSuc10,Suchardetal10,Wenetal11,Altetal13,Chenetal13,Leeetal16,FajLia17}.  However, a general extension of the EM algorithm that is tuned for applications in grid computing environments and has theoretical convergence guarantees remains unknown. The main challenge here is to retain the generality of E step while minimizing the computational bottleneck due to extensive communication among processes.  

The DEM algorithm is designed for extending EM-type algorithms, which work on a single machine, to the distributed setting with minimal modifications. DEM first reserves $(K + M)$ processes for computations, where $M \ll K$ and $K$ and $M$ are the number of workers and managers, respectively. The samples in the full data are randomly partitioned into $K$ disjoint subsets that are stored on the {workers}. The {managers} manage the communications among workers and track the progress of DEM, including maintaining the latest copies of E step results received from all the worker processes. The E step of DEM algorithm consists of performing the local E step in parallel on the worker processes, and the result of every local E step is communicated to the managers. For a $\gamma \in (0, 1]$, the managers receive 
results from a $\gamma$-fraction of the workers and perform the M step using an objective that depends on the $\gamma$-fraction of new E step results and the ($1-\gamma$)-fraction of old E step results. The managers stop the DEM iterations when the likelihood has reached a local mode. The local E step on every subset is fast and free of any memory limitations. The asynchronous M step minimizes computational bottlenecks due to extensive communication among workers and managers. 

The DEM algorithm is a generalization of the EM algorithm in that it reduces to a classical but distributed EM algorithm if $\gamma=1$. DEM differs from the previous EM extensions in its distributed and fractional updates. {Many EM-based model fitting methods in robust statistics also use only a fraction of the data; see for example \citet{Neyetal07}. This idea differs from that of DEM which uses results from every subset for each iteration but only a $\gamma$-fraction of these results are new.} DEM's sequence of parameter estimates converges to a local mode under the theoretical setup of the classical EM. This is a major advantage relative to existing distributed EM extensions that are restricted to a particular class of models or likelihoods. Existing distributed EM extensions are special cases of the DEM depending on $\gamma$. For example, the distributed EM algorithms for mixture models are a special case of DEM if $\gamma = 1$; and DEM reduces to the Incremental EM (IEM) \citep{NeaHin98} if $\gamma = 1/n$, where $n$ is the sample size.  Our numerical experiments show that DEM is also easy to implement on cluster of computers using a non-distributed implementation of an EM-type algorithm and the message passing interface (MPI) \citep{Gabetal04}. 

The run-time efficiency of DEM relative to its non-distributed version depends on $K$ and $\gamma$. DEM with $\gamma \in (0, 1)$ requires more iterations to reach a local mode than its non-distributed version because it uses only a $\gamma$-fraction of the full data for each iteration. If $\gamma$ is close to 1, then the number of iterations required for convergence in DEM is very similar to that of the non-distributed version but the communication overhead is large; if $\gamma$ is close to 0, then the communication overhead is small but the number of iterations required for convergence is relatively large.  If $K$ is chosen to be large enough so that the local E steps finish quickly and communication cost is of the order $O(1)$, then DEM algorithm can be faster that its non-distributed version for a broad range of $\gamma$. Empirically, DEM is more than two times faster than its non-distributed version for $\gamma$ values around $0.5$, which achieves the optimal balance
between the decreased computational burden due to efficient local E steps and increased number of iterations required for convergence. 

\section{Motivating example: MovieLens ratings data}
\label{sec:motivating-example}

MovieLens data are one of the largest publicly available movie ratings data (\url{http://grouplens.org}). The database contains 10 million ratings for about 11 thousand movies by about 72 thousand users of the online movie recommender service MovieLens, where a user has rated multiple movies. An observation in the database includes movie rating from 0.5 to 5 in 0.5 increments, time of the rating, and 18 genres to which the movie belongs. A user rating can be predicted using a linear mixed-effects model with movie genres as fixed and random covariates. A variety of efficient EM-type algorithms are available for fitting mixed-effects models; however, they are slow due the time consuming E step. Motivated by similar problems in using state-of-the-art lme4 R package \citep{Batetal13}, \citet{Per17} proposed a new approach for fitting such models using the method of moments. 

We extended an ECME algorithm in \citet{Van00} using the DEM algorithm to achieve two time speed ups in fitting linear mixed-effects model to the MovieLens data with $K=20$, and $m=1$. We reserved 20 worker processes and a manager process on a cluster, randomly split the users into 20 disjoint subsets, and stored their data on the workers. The E step  was performed in parallel on the 20 workers, and the manager performed the M step. DEM was implemented in R \citep{R16} using the Rmpi package \citep{Yu02} and van Dyk's ECME algorithm. Our real data analysis based on the examples considered in \citet{Per17} showed that DEM with $\gamma = 0.3, 0.5,0.7$ matches the accuracy of van Dyk's ECME in parameter estimation while being significantly faster for all three values of $\gamma$; see Section \ref{expr}. The major advantage of DEM was its generality in that it scaled the ECME algorithm to massive data settings using its non-distributed implementation and MPI.

\section{Basic setup of DEM algorithm}
\label{sec:fram}

Consider a general framework for implementing any iterative statistical algorithm when the data are stored on multiple processes in a distributed or grid computing environment. Assume that the distributed environment has a collection of $K$ processes that store the $K$ disjoint data subsets and that are responsible for computations for each iteration of the statistical algorithm. At the $t$-th iteration of the algorithm, process $k$ computes quantities $h_{t,k}= H(\theta_t, S_k)$ determined by the assigned data subset $S_k$ and the current state $\theta_t$ of a common quantity $\theta$ shared by all the data subsets. In the context of EM algorithm, $H(\theta_t, S_k)$ is the conditional expectation of the sufficient statistics given $S_k$ and the current estimate $\theta_t$ of the parameter $\theta$. The common or population quantity $\theta$ is updated by
\begin{equation}\label{eq:ipc-update}
\theta_{(t+1)} = G(\theta_t; h_{t,1}, \ldots, h_{t,k}, \ldots, h_{t, K}),
\end{equation}
where $h_{t,k} = H(\theta_t, S_k)$, $k=1, \ldots, K$. This provides a general setup for any distributed iterative statistical algorithm, including the DEM.

Consider the Inter-Process Communication (IPC) scheme for implementing distributed statistical algorithms that can be described using (\ref{eq:ipc-update}). For process $k=1, \ldots, K$,
\begin{algor}\label{alg:ipc-naive}
Starting with $\theta_0$, iterate between the following two steps for $t=0, \ldots, \infty$.
\begin{enumerate}
\item[(a)] Compute $h_{t,k}=H(\theta_t, S_k)$ and send $h_{t,k}$ to all the other processes.
\item[(b)] Upon receiving all $h_{t, k}$ values for $k = 1, \ldots, K$, evaluate (\ref{eq:ipc-update}) to obtain $\theta_{(t+1)}$.
\end{enumerate}
\end{algor}
Although conceptually simple, such a generic scheme fails in a distributed setting where communication between processes is time consuming or even unreliable. \citet{Chenetal13} proposed an EM extension for mixture models based on this scheme and employed single-program-multiple-data paralellization technique to achieve efficiency; however, generalizations of this approach to other EM-type algorithms are unclear.

Statistical thinking in terms of \emph{imputation-and-analysis} steps for iterative algorithms, such as EM, make it appealing to consider distributed computing environments with few \emph{manager processes} for ``analysis'' and a large number of \emph{worker processes} for ``imputation.'' Accordingly, the version of the Scheme \ref{alg:ipc-naive} in the distributed environments with one manager process and $K$ worker processes is given as follows. 
\begin{algor}\label{alg:ms-ipc-update}
This scheme consists of two iterative sub-schemes, one for the workers  processes and another for the manager  process.
The {\bf manager} process  starts with $\theta_0$ and iterates between the following two steps for $t=0, \ldots, \infty$.
\begin{enumerate}
\item[(a)] Send $\theta_{t}$ to all the worker processes.
\item[(b)] Wait to receive $h_{t,k}$ from all $k=1,...,K$, and use (\ref{eq:ipc-update}) to obtain $\theta_{(t+1)}$.
\end{enumerate}
The {\bf worker} process $k$ iterates between the following two steps for $t=0, \ldots,\infty$.
\begin{enumerate}
\item[(a)] Wait to receive $\theta_{t}$ from the manager process.
\item[(b)] Compute $h_{t,k} = H(\theta_t, S_k)$ and send $h_{t,k}$ to the manager process.
\end{enumerate}
\end{algor}
Scheme \ref{alg:ms-ipc-update} avoids the expensive communication overheads of Scheme \ref{alg:ipc-naive}, but algorithms implemented using this scheme can be dramatically slow if the computational burden on a few worker processes is large. Unfortunately, such events are typical in distributed environments where limited computational resources are shared by many processes. 

Scheme 2 is adopted by most distributed implementations of EM. \citet{ZhoLanSuc10} and \citet{Suchardetal10} have used a GPU framework, where threads are worker processes and no manager is required because the computations are performed on a single node. Efficiency is maintained by exploiting the GPU architecture and avoiding data copying between the GPUs and the host machine. \citet{Leeetal16} use a multi-threading framework that does not require data copying and communication. These approaches are best suited for computations on shared-memory architectures. The E step in all these methods uses data from all the worker threads, which can be slow if one thread has a large computational burden.

An asynchronous modification of Scheme \ref{alg:ms-ipc-update} allows the manager to update once it has received at least a pre-specified $\gamma$-fraction of updated $h_{t,k}$'s. The manager updates $\theta$ with  $h_{t,k}$'s fixed at their most recent values. A simplified version of this scheme is as follows. 
\begin{algor}\label{alg:aipc}
Given $\gamma \in (0, 1)$ and denoting the most recent value of $h_{t,k}$ as $h_{t,k}^{*}$, the {\bf manager} process starts with $\theta_0$ and $h^*_{0, 1}, \ldots, h_{0, K}^*$, and iterates between the following two steps for $t=0, ..., \infty$.
\begin{enumerate}
\item[(a)] Wait until having received at least a  $\gamma$-fraction of $h_{t, k}$'s.  
\item[(b)] Compute
\[
\theta_{(t+1)} = G(\theta_{t}; h_{t, 1}^*, ..., h_{t, K}^*)
\]
and send $\theta_{(t+1)}$ to all the worker processes.

\end{enumerate}
The {\bf worker} process $k$ iterates between the following two steps for $t=0, \ldots,\infty$.
\begin{enumerate}
\item[(a)] Wait to receive $\theta_{t}$ from the manager process.
\item[(b)] Compute $h_{t,k}=H(\theta_t, S_k)$ and send $h_{t,k}$ to the manager process.
\end{enumerate}
\end{algor}
Scheme \ref{alg:aipc} has many desirable properties for implementing iterative statistical algorithms, such as EM, in distributed environments. The DEM algorithm provides a general framework for implementing any EM-type algorithm in distributed computing environments using Scheme \ref{alg:aipc}. The formal definition of the DEM algorithm is given in Section \ref{sec:definition-dem}.

\section{Basic theory of DEM}
\label{sec:conv}

\subsection{Notation and background}
\label{sec:notation-background}

Consider a general EM algorithm setup with $Z = \{z_1, \ldots, z_n\}$ representing the full data consisting of $n$ samples. The samples in full data are randomly partitioned into $K$ disjoint subsets. Represent the data in subset $k$ as 
$Z_k = \{z_{k1}, \ldots, z_{k n_k}\}$  $(k=1, \ldots, K)$, where $z_{kj} = z_i$ for some $i \in \{1, \ldots, n\}$ and every $j=1, \ldots, n_k$, so the full data $Z = \{Z_1, \ldots, Z_K\}$. The data subsets $Z_1, \ldots, Z_K$ are stored separately on $K$ workers. Let $g(Z_{1:K} \mid \theta) = \prod_{k=1}^K g(Z_k \mid \theta)$ be the density of $Z_{1:K}$ based on its probability model parametrized by $\theta$ lying in some space $\Theta$, where $Z_{1:K}$ is a shorthand for the sequence $Z_1, \ldots, Z_K$. The log likelihood of $\theta$ given the observed data is
\begin{align}
  \Lcal(\theta) = \sum_{k=1}^K \log g(Z_k \mid \theta) \equiv \sum_{k=1}^K \Lcal_k(\theta), \quad \Lcal_k(\theta) = \log g(Z_k \mid \theta), \label{log-lik}
\end{align}
where $\Lcal_k(\theta)$ represents the contribution of the data on process $k$ to the log likelihood ($k=1, \ldots, K$). The maximum likelihood estimate (MLE) of $\theta$ in the parameter space $\Theta$ is 
\begin{align}
  \hat \theta = \underset{\theta \in \Theta} {\argmax} \,  \Lcal(\theta) = \underset{\theta \in \Theta} {\argmax} \, \sum_{i=1}^K \log g(Z_i \mid \theta) = \underset{\theta \in \Theta} {\argmax} \, \sum_{i=1}^K \Lcal_k(\theta). \label{max-lik}
\end{align}
Finding $\hat \theta$ by direct maximization in (\ref{max-lik}) is difficult in many statistical applications. EM algorithm simplifies this problem by augmenting ``missing'' data $Y_k$ to $Z_k$, yielding  complete-data $(Y_k, Z_k)$ for subset $k$ ($k=1, \ldots, K$). The joint density of the complete-data $f(Y_{1:K}, Z_{1:K} \mid \theta) = \prod_{k=1}^K f\{(Y_k, Z_k \mid \theta )\}$ still depends on $\theta$ and marginalizing missing-data from the joint yields the density of observed data  
\begin{align}
  \label{g-f}
  g(Z_{1:K} \mid \theta) = \int f(Y_{1:K}, Z_{1:K} \mid \theta) d Y_{1:K} = \prod_{k=1}^K g(Z_{k} \mid \theta), \; g(Z_{k} \mid \theta) = \int f(Y_{k}, Z_{k} \mid \theta) d Y_{k}.
\end{align}

The EM algorithm maximizes $\Lcal(\theta)$ (\ref{log-lik}) by iteratively maximizing a modified form of $\log f(Y_{1:K}, Z_{1:K} | \theta)$. Let $\theta_{t}$ represent the estimate of $\theta$ at the end of $t$-th iteration of EM. The E step at the $(t+1)$-th iteration replaces $\log f(Y_{1:K}, Z_{1:K} \mid \theta)$ by its conditional expectation with respect to the conditional density of $Y_{1:K}$ given $Z_{1:K}$ with parameter $\theta_t$, denoted as $h(Y_{1:K} \mid Z_{1:K}, \theta_t)$, to obtain
\begin{align}
  Q(\theta \mid \theta_{t}) = \EE_Y \left\{ \log f(Y_{1:K}, Z_{1:K} \mid \theta) \mid  Z_{1:K} \right\} 
  = \sum_{k=1}^K Q_k(\theta \mid \theta_{t}),   \label{q-fun}
\end{align}
where $Q_k(\theta \mid \theta_{t}) = \EE_Y \left\{ \log f (Y_k, Z_k \mid \theta ) \mid Z_k  \right\}$, $\EE_Y$ represents expectation with respect to $h(Y_{1:K} \mid Z_{1:K}, \theta_t)$, and $ Q_k(\theta \mid \theta_{t})$ represents the contribution of worker $k$ to $Q(\theta \mid \theta_t)$. The M step finds the $(t+1)$-th update of $\theta$ as 
\begin{align}
  \theta_{(t+1)} = \underset{\theta \in \Theta} {\argmax} \,  Q(\theta \mid \theta_{t}) = \underset{\theta \in \Theta} {\argmax} \sum_{k=1}^K Q_k(\theta \mid \theta_{t}). \label{em-its}
\end{align}
Let $H(\theta \mid \theta_{t}) = \EE_Y \left\{ \log h(Y_{1:K} | Z_{1:K}, \theta) | Z_{1:K} \right\}$. Then, at the $t$-th iteration,
\begin{align}
  \Lcal(\theta) =  Q(\theta \mid \theta_{t}) -  H(\theta \mid \theta_{t}) = \sum_{k=1}^K Q_k(\theta \mid \theta_{t}) -  \sum_{k=1}^K H_k(\theta \mid \theta_{t}),  \label{q-h}
\end{align}
where $H_k(\theta \mid \theta_{t})$ represents the contribution of worker $k$ to $H(\theta \mid \theta_{t})$. As a function of $\theta$, $H(\theta \mid  \theta_{t})$ is maximized at $\theta_t$, so $\Lcal(\theta_{t+1}) \ge \Lcal(\theta_{t})$ for $\theta_{(t+1)}$ defined in \eqref{em-its}; see Theorem 1  of \citet{DemLaiRub77}. Any version of the EM algorithm that ensures the ascent of $\Lcal(\theta)$ is called a generalized EM (GEM) algorithm \citep{DemLaiRub77}.

\citet{NeaHin98} present an alternative interpretation of the EM algorithm that greatly simplifies the theory and results related to DEM. They show that the E and M steps of any GEM algorithm respectively maximize a common functional $F(\widetilde p, \theta)$, where $\widetilde p$ represents an unknown density of $Y_{1:K}$ parametrized by $\phi$. The E step estimates $\widetilde p$ by maximizing the objective functional 
\begin{align*}
  F(\widetilde p, \theta) &= \widetilde \EE_Y \left\{ \log f(Y_{1:K}, Z_{1:K} \mid \theta) \right\} - \widetilde \EE \left\{ \log \widetilde p (Y_{1:K}) \right\}, 
\end{align*}
where $\widetilde \EE_Y$ denotes expectation with respect to the density $\widetilde p(Y_{1:K} \mid \phi)$. After some algebra, this reduces to 
\begin{align}
  F(\widetilde p, \theta) &= -\text{KL} \left\{ \widetilde p(Y_{1:K} \mid \phi), h(Y_{1:K} \mid Z_{1:K}, \theta) \right\} + \log g(Z_{1:K} \mid \theta), \nonumber\\ 
&=  \sum_{k=1}^K \left[ -\text{KL} \left\{ \widetilde p(Y_{k} | \phi),  h(Y_{k} | Z_{k}, \theta) \right\} + \Lcal_k(\theta) \right], \label{nea-hin-2}
\end{align}
where $\text{KL}(\widetilde p, h) = \int \log (\widetilde p / h) \widetilde p \, d y$ is the Kullback-Liebler (KL) divergence between $\widetilde p$ and $h$. Theorem 1 of \citet{NeaHin98} shows that the E step in $(t+1)$-th iteration maximizes  $F(\widetilde p, \theta)$ in (\ref{nea-hin-2})  by setting $\widetilde p = \prod_{k=1}^K \widetilde p_{(t+1), k} \equiv \widetilde p_{(t+1)}$  for a fixed  $\theta_{t}$, where $ \widetilde p_{(t+1), k} = h(Y_k \mid Z_k, \theta_t)$ ($k=1, \ldots, K$). The M step then maximizes $F(\widetilde p_{(t+1)}, \theta)$ with respect $\theta$ for a fixed $\widetilde p_{(t+1)}$. These two steps are repeated until convergence to the stationary point $F(\hat{\widetilde p}, \hat \theta)$.  Theorem 2 of \citet{NeaHin98} shows that if $F$ has a global or local maximum at $\hat {\widetilde p}$ and $\hat \theta$, then $\Lcal(\theta)$ has a global or local maximum at $\hat \theta$. Based on this observation, \citet{NeaHin98} propose the IEM algorithm that cyclically updates $\widetilde p$ and $\theta$ separately based on the $i$-th sample for $i=1, \ldots, n$; see equations (7), (8), and (9) in \citet{NeaHin98}. 

\subsection{E and M steps of DEM}
\label{sec:definition-dem}

The DEM algorithm is an asynchronous and distributed generalization of the EM algorithm based on Scheme \ref{alg:aipc}.  In any iteration of the DEM algorithm, the managers for the DEM algorithm maintain a copy of $Q_k$, $H_k$, and $\Lcal_k$ based on the last communication with worker $k$ ($k=1, \ldots, K$). The worker $k$ performs its local E step of the DEM algorithm using its data subset $Z_k$, calculates $Q_k$, and returns its $Q_k$ to the managers. The M step is performed by the manager machines when they have received $Q_k$'s from $N$ processes such that $N/K \ge \gamma$. After the M step, the managers send the updated $\theta$ to all the processes for the next iteration of DEM algorithm. This process is repeated until convergence to the local mode $\hat \theta$. If $\gamma = 1$, then DEM follows the synchronous Scheme \ref{alg:ms-ipc-update} and reduces to the classical but distributed EM. 

The DEM iterations are defined using the notation introduced in the previous section. Denote $Q_k(\theta \mid  \theta_{t_k})$, $H_k(\theta \mid \theta_{t_k})$, and $\Lcal_k(\theta)$ respectively as the $Q_k$, $H_k$, $\Lcal_k$ functions and $\theta_{t_k}$ as the latest copy of $\theta$ maintained by the managers for the worker $k$ ($k=1, \ldots, K$) at the $t$-th iteration. At the start of $(t + 1)$-th iteration, the managers send the current parameter estimate $\theta_t$ to all the processes and DEM proceeds as follows.
\begin{description}
\item[E step:] For $k=1, \ldots, K$, worker $k$ computes its $Q_k(\theta \mid \theta_t) =  \EE_Y \left\{ \log f(Y_k, Z_k \mid \theta) \mid Z_k \right\} $ and returns the $Q_k(\theta \mid \theta_t)$ to the managers.
\item[M step:] The managers wait until they have received $Q_k(\theta \mid \theta_t)$'s from $N$ workers, where $N$ is such that $N/K \ge \gamma$. Once the managers are done with receiving, they calculate the $(t+1)$-th update for $\theta$ as
 \begin{align}
    \theta_{(t+1)} =  \underset{\theta \in \Theta} {\argmax} \, \bigg\{\sum_{k \in U_{(t+1)}} Q_k(\theta \mid \theta_{t}) + \sum_{k \in U_{(t+1)}^c} Q_k(\theta \mid \theta_{t_k}) \bigg\},  \label{m-step}
  \end{align}
where $U_{(t+1)}$ contains the indices of processes that returned their $Q_k$'s to the managers and $U_{(t+1)}^c = \{ 1, \ldots, K\} \backslash U_{(t+1)}$. The managers send $\theta_{(t+1)}$ to all the workers for the next iteration, including the workers that did not return their $Q_k$'s to the manager. 
\end{description}
Later we assume that every worker returns its $Q$-function to the managers infinitely often. Under this assumption, if we relabel the processes that returned their $Q_k$'s as $k=1, \ldots, N$ and the remaining processes as $k=(N+1), \ldots, K$, then the M step in (\ref{m-step}) reduces to
\begin{align}
  \theta_{(t+1)} =  \underset{\theta \in \Theta} {\argmax} \, \bigg\{ \sum_{k = 1}^N Q_k(\theta \mid \theta_{t}) + \sum_{k = N+1}^K Q_k(\theta \mid \theta_{t_k}) \bigg\} \equiv  \underset{\theta \in \Theta} {\argmax} \, Q(\theta \mid \theta_{t}, \theta_{t_{N+1}}, \ldots, \theta_{t_{K}}).  \label{m-step-lbl}
\end{align}

The E and M steps of DEM are repeated until $\Lcal(\theta_t)$ sequence converges. Theorem \ref{dem-gem} proves that the sequence $\Lcal(\theta_t)$ indeed has a stationary point. 
\begin{theorem} \label{dem-gem} 
  The sequence $F(\widetilde p_t, \theta_t)$ does not decrease in  DEM; that is, $F(\widetilde p_t, \theta_t) \leq F(\widetilde p_{t+1}, \theta_{t+1})$,  $t \geq 0$. If $F(\widetilde p, \theta)$ is bounded above, then  $F(\widetilde p_t, \theta_t) \rightarrow \hat F$ for some $\hat F$. In particular, $\Lcal(\theta_{t}) \rightarrow \hat \Lcal$ for some $\hat \Lcal$.
\end{theorem}
The proof is in the supplementary material along with other proofs. This is a desirable but a weaker result in that DEM fails to maintain the monotone ascent of the likelihood sequence. While DEM is not a GEM, Theorem \ref{dem-gem} implies that there exists a likelihood subsequence $\Lcal(\theta_{t_j})$ that maintains the monotone ascent of the likelihood in that $\Lcal(\theta_{t_{j + 1}}) \ge \Lcal(\theta_{t_j})$, $j \geq 0$. If we define a weak-GEM to be an EM-type algorithm that maintains the monotone ascent of a likelihood subsequence, then DEM is a weak-GEM, and in the same fashion as online EMs and the IEM.

The M step in (\ref{m-step-lbl}) can be modified based on any efficient extension of the classical EM, such as ECM, ECME, PX-EM. The proof of Theorem \ref{dem-gem} implies that DEM maintains the monotone ascent for $F(\tilde p, \theta)$ in \eqref{nea-hin-2}. The $Q_k$ functions in (\ref{m-step-lbl}) can be replaced by any other function such that $\theta_{t+1}$ does not decrease $F(\tilde p_{t+1}, \theta)$. For example, the $(t+1)$-th update for $\theta$ defined as
\begin{align}  
  \theta_{(t+1)} =  \underset{\theta \in \Theta} {\argmax} \, \bigg\{ \sum_{k = 1}^N Q_k(\theta \mid \theta_{t}) + \sum_{k = N+1}^K F_k(\tilde p_{t+1}, \theta) \bigg\}                    , \label{ecme-step}
\end{align}
which guarantees $F(\tilde p_{t+1}, \theta_{t+1}) \ge F(\tilde p_t, \theta_t)$, $t \geq 0$. We use this idea in our simulated and real data analyses to implement distributed extensions of ECME algorithm. The IEM algorithm of \citet{NeaHin98} is obtained by fixing $K = n$ and by modifying (\ref{m-step}) as $\theta_{(t+1)} =  \underset{\theta \in \Theta} {\argmax} \, Q_k(\theta \mid \theta_{t})$ for a $k \in \{ 1, \ldots, n\} $, where $k$ can be chosen randomly or in a specific order. This implies that IEM is a DEM if the $n$ samples are treated as subsets and $\gamma = 1/n$.

\citet{Wu83} shows that more regularity conditions are needed to guarantee that the DEM sequence $\theta_t$ converges to $\hat \theta$, a local mode or stationary point. Proving a similar result for DEM is difficult because DEM is not a GEM and the $\theta_t$ sequence in DEM depends on multiple previous iterates. We modify the arguments in \citet{Wu83} using $F(\tilde p, \theta)$ as the ascent function to obtain the global convergence result for the DEM sequence. Define $\Pi = \{\tilde p : \text{KL} \{\tilde p(y), h(y \mid Z, \theta)\} < \infty \text{ for every } \theta,\,  \int \log \{\tilde p(y) \} \tilde p(y) dy < \infty \}$. Our setup has the following assumptions:
\begin{enumerate}    
\item  \label{a1} $\Theta$ is a subset in the $P$-dimensional Euclidean space $\RR^P$.
\item \label{a2} The set $\Pi_{\theta_0} \otimes \Theta_{\theta_0} = \{(\widetilde p, \theta) \in \Pi \otimes \Theta: \, F(\widetilde p, \theta) \ge F(\tilde p_0, \theta_0) \}$ is compact for any starting point of the ($\tilde p_t$, $\theta_t$) sequence, denoted as $(\tilde p_0, \theta_0)$, that satisfies $ \Lcal(\theta_0) > -\infty$ and $\tilde p_0 = \prod_{k=1}^K h(Y_k \mid Z_k, \theta_0)$.   
\item  \label{a3} $F(\tilde p, \theta)$ is continuous in $\Pi \otimes \Theta$ and differentiable in the interior of $\Pi \otimes \Theta$.
\item  \label{a4} $\Pi_{\theta_0} \otimes \Theta_{\theta_0}$ is in the interior of $\Pi \otimes \Theta$ for any $\theta_0 \in \Theta$.  
\item  \label{a5} The first order differential ${\partial Q(\theta \mid \theta_{t}, \theta_{t_{(N+1)}}, \ldots, \theta_{t_{K}})} / {\partial \theta}$ is continuous in $(\theta, \theta_t, \theta_{t_{(N+1)}}, \ldots, \theta_{t_{K}})$.    
\item  \label{a6} Worker $k$ returns $Q_k$ to the manager infinitely often for $t=0, \ldots, \infty$ and $k = 1, \ldots, K$.
\end{enumerate}
Assumptions \ref{a1}--\ref{a5} follow from \citet{Wu83}. Assumption \ref{a2} implies that $(\tilde p_t, \theta_t)$ sequence is bounded. Assumptions \ref{a1}--\ref{a3} and the definition of $\Pi$ imply that $F(\tilde p, \theta)$ is bounded above for any $\theta \in \Theta$. Assumption \ref{a4} guarantees the existence of  derivatives of $F(\tilde p, \theta)$, $\text{KL} \{\tilde p, h(Y \mid Z, \theta)\}$, $\Lcal(\theta)$ at $(\tilde p_t, \theta_t)$. Assumption \ref{a5} is used to show that $F(\tilde p_t, \theta_t)$  sequence  converges monotonically to $F(\hat{ \widetilde p}, \hat \theta)$ for a stationary point $(\hat{ \widetilde{p}}, \hat \theta)$ as in Theorem 2 of \citet{Wu83}. Assumption \ref{a6} ensures that $\Lcal(\theta)$ uses the full data as $t \rightarrow \infty$ and is used later in deriving the matrix rate of convergence of the DEM in the next section. These assumptions hold for the linear mixed-effects model used in our simulated and real data analysis. 

The next theorem describes the convergence of $(\tilde p_t, \theta_t)$ sequence, which implies the convergence of $\theta_t$ sequence. Let $\Scal$ be the set of stationary points and $\Mcal$ be the set of local maxima in the interior of $\Pi \otimes \Theta$. For a given $F$, define the sets
$\Scal( F) = \{(\tilde p, \theta) \in \Scal : F(\tilde p, \theta) =  F\}$ and $\Mcal( F) = \{(\tilde p, \theta) \in \Mcal : F(\tilde p, \theta) =  F\}$. 
\begin{theorem}\label{conv-dem}
  Suppose Assumptions \ref{a1}--\ref{a6} hold. Then,
  \begin{enumerate}
  \item if $\Scal(\hat F)$ (resp. $\Mcal(\hat F)$) = $\{(\hat {\widetilde p}, \hat \theta)\}$, where $\hat F$ is  the limit of $F(\tilde p_t, \theta_t)$ sequence, then $(\tilde p_t, \theta_t) \rightarrow (\hat{ \widetilde p}, \hat \theta)$, implying that $\theta_t \rightarrow \hat \theta$, where $\hat \theta$ is a stationary point (resp.  local maximum) of $\Lcal(\theta)$; and
  \item if $\Scal(\hat F)$ (resp. $\Mcal(\hat F)$) is discrete and $\|(\tilde p_{t+1}, \theta_{t+1}) - (\tilde p_t, \theta_t)\|_{\Pi \otimes \Theta} \rightarrow 0$ as $t \rightarrow \infty$, where $\| \cdot \|_{\Pi \otimes \Theta}$ is a norm on $\Pi \otimes \Theta$, then $(\tilde p_t, \theta_t) \rightarrow (\hat{\widetilde {p}}, \hat \theta)$  
  for some $(\hat{\tilde {p}}, \hat \theta)$ in $\Scal(\hat F) \cup \Mcal(\hat F)$, implying that $\theta_t \rightarrow \hat \theta$,      where $\hat \theta$ is a stationary point (resp.  local maximum) of $\Lcal(\theta)$.
  \end{enumerate}  
\end{theorem}
This theorem strengthens Theorem 2 in \citet{NeaHin98} because it describes the convergence of $(\tilde p_t, \theta_t)$ sequence, which is not implied by the convergence of $F(\tilde p_t, \theta_t)$ sequence. 
There are exceptional cases due to uneven load sharing on the workers where Assumption \ref{a6} can be violated.  Theorem \ref{conv-dem}  is inapplicable in those cases.  

\subsection{Matrix rate of convergence of DEM}
\label{sec:rate-convergence-dem}

We compare the matrix rate of convergence of DEM sequence $\{\theta_t, t \geq 0\}$ to its non-distributed version from the managers' perspective. Distributed and asynchronous computations are an important component of DEM, but for simplicity we use the tools developed in \citet{DemLaiRub77}, \citet{Men94}, and \cite{LiuRubWu98} for deriving the matrix rate of convergence of DEM and assume that the cost of communication among workers and managers is negligible. We show that the difference between the rates of convergence of DEM and the classical EM depends on the observed information and complete-data information matrices calculated using the $\gamma$-fraction and $(1-\gamma)$-fraction of the full data.

Consider an EM sequence $\{\theta^E_t, t \geq 0\}$. Each EM iteration defines a mapping $M_{\ema}$ so that $\theta^E_{(t+1)} = M_{\ema}(\theta^E_t)$. If $\hat \theta^E$ is a fixed point of the $\theta^E_t$ sequence, then a Taylor expansion at $\hat \theta^E$ gives $\theta^E_{(t+1)} = D M_{\ema} (\theta^E_t - \hat \theta^E) + o(\| \theta^E_t - \hat \theta^E\|)$, where $D M_{\ema}$ is the gradient of map $M_{\ema}$, $\tfrac{\partial M_{\ema}}{\partial \theta}$, evaluated at $\hat \theta^E$ and $o(t) / t \rightarrow 0$ as $t \rightarrow 0$. The matrices $D M_{\ema}$ and $S_{\ema} = I - DM_{\text{EM}}$, where $I$ is an identity matrix of appropriate dimension, are called the matrix rate of convergence and speed matrix  of EM, respectively. Let $\lambda_{\min}(A)$ and $\lambda_{\max}(A)$ represent the minimum and maximum singular values of a matrix $A$. Then, $\lambda_{\max}(DM_{\text{EM}})$ and $\lambda_{\min}(S_{\text{EM}}) = 1 - \lambda_{\max}(DM_{\text{EM}})$  respectively are the global rate and global speed of convergence of EM. \citet{DemLaiRub77} show that 
\begin{align}
  \label{eq:em1}
  S_{\ema} =  i_{\text{com}}^{-1} i_{\text{obs}},\; i_{\text{obs}} = - \tfrac{\partial^2 \log g(Z_{1:K} \mid \theta)} {\partial \theta \cdot \partial \theta^T} \big|_{\theta = \hat \theta^E}, \; i_{\text{com}} = - \EE_Y \left\{ \tfrac{\partial^2 \log f(Y_{1:K}, Z_{1:K} \mid \theta)} {\partial \theta \cdot \partial \theta^T} \mid Z_{1:K}, \theta \right\}  \big|_{\theta = \hat \theta^E}, 
\end{align}
where $i_{\text{obs}}$ and $i_{\text{com}}$ are the observed-data and complete-data information matrices. 

These techniques require modification before their application to deriving the rate of convergence of DEM. Assume that $N/K=\gamma$ and consider the DEM sequence $\{(\theta_{t_1}, \ldots, \theta_{t_K}), t \geq 0\}$ at the managers for estimating $\theta$, where $\theta_{t_k}$ corresponds to the value of $\theta_t$ in $Q_k(\theta \mid \theta_{t})$ at the $t$th iteration. If $k \in U_t$, then $\theta_{t_k}=\theta_t$; otherwise, $\theta_{t_k}=\theta_{s}$ for some $s < t$ ($k=1, \ldots, K$). Since $U_t$s are independent random sets of indices, the gradient of DEM mapping $M_{\dem}$ such that $\theta_{(t+1)} = M_{\dem}(\theta_t)$ is not well-defined; therefore, we choose a subsequence of $\{(\theta_{t_{j_1}}, \ldots, \theta_{t_{j_K}}), j \geq 0\}$ corresponding to those $t_j$s such that M step in \eqref{m-step} has $U_{t_j} = \{1, \ldots, N\}$ and the objective is \eqref{m-step-lbl}. This implies that $\theta_{t_{j_1}}, \ldots, \theta_{t_{j_N}}$ are equal. If $\theta_{t_{j_1}} = \cdots = \theta_{t_{j_N}} = \theta_{t_{j}}$, then we represent this DEM subsequence as $\theta^D_j = (\theta_{t_j}, \theta_{t_{j_{N+1}}}, \ldots, \theta_{t_{j_K}})$ and $\theta^D_j \in \RR^{P (K - N +1)}$. Borel-Cantelli lemma implies the existence of $\{\theta^D_j, j \geq 0\}$ with positive probability using Assumption 6 in Theorem \ref{conv-dem} and the independence of $U_t$s for $t \geq 0$. 

Our rate of convergence results from the managers' perspective are derived by focusing on $\{\theta_{t_j}, j \geq 0\}$, the first $P$ coordinates of the DEM subsequence $\{\theta^D_j = (\theta_{t_j}, \theta_{t_{j_{N+1}}}, \ldots, \theta_{t_{j_K}}), j \geq 0\}$.  The subsequence $\{\theta_{t_j}, j \geq 0\}$ and $K-N$ subsequences $\{\theta_{t_{j_i}}, j \geq 0\}$ ($i=N+1, \ldots, K$) converge because they are subsequences of the convergent DEM sequence $\{\theta_t, t \geq 0\}$; see Theorem \ref{conv-dem}. Since $\{\theta^D_j, j \geq 0\}$ is a vector of convergent sequences, $\hat{ \theta}^D = (\hat \theta, \hat \theta, \ldots, \hat \theta) \in \RR^{P (K - N +1)}$ is one of its fixed point. Let $M_{D}$ be a mapping such that $\theta_{(j+1)}^D = M_D(\theta^D_j)$ and $\theta^D_{(j+1)} = D M_D (\theta^D_j - \hat \theta^D) + o(\| \theta^D_j - \hat \theta^D\|)$. Represent $\theta^D_j$  as $(\theta_{j}, \psi_{j})$, where $\psi_{j} = (\theta_{t_{j_{N+1}}}, \ldots, \theta_{t_{j_K}})\in \RR^{P(K-N)}$. Define the observed-data and complete-data information matrices for $\{\theta_{j}, j \geq 0 \}$ and $\{\psi_{j}, j \geq 0\}$ sequences as
\begin{align}
  \label{eq:em2}
  i_{\text{obs}, \theta} &= - \tfrac{\partial^2 \log g(Z_{1:N} \mid \theta)} {\partial \theta \cdot \partial \theta^T} \big|_{\theta = \hat \theta}, \quad
  i_{\text{obs}, \psi} = - \text{bdiag} \left\{ \tfrac{\partial^2 \log g(Z_{N+1} \mid \theta)}{\partial \theta \cdot \partial \theta^T} \big|_{\theta = \hat \theta}, \ldots, \tfrac{\partial^2 \log g(Z_{K}   \mid \theta)}{\partial \theta \cdot \partial \theta^T}\big|_{\theta = \hat \theta}\right\}, \nonumber \\
  i_{\text{com}, \theta} &= - \EE_Y \big\{ \tfrac{\partial^2 \log f(Y_{1:N}, Z_{1:N} \mid \theta)}{\partial \theta \cdot \partial \theta^T} \mid  Z_{1:N}, \theta \big\} \big|_{\theta = \hat \theta},\\
  i_{\text{com}, \psi} &= - \text{bdiag} \big[\EE_Y \big\{ \tfrac{\partial^2 \log f(Y_{(N+1)}, Z_{(N+1)} \mid \theta)}{\partial \theta \cdot \partial \theta^T} \mid Z_{(N+1)}, \theta \big\} \big|_{\theta = \hat \theta},  \ldots, \EE_Y \big\{ \tfrac{\partial^2 \log f(Y_{K}, Z_{K} \mid \theta)}{\partial \theta \cdot \partial \theta^T} \mid Z_{K}, \theta \big\} \big|_{\theta = \hat \theta} \big], \nonumber
\end{align}
where $\text{bdiag}(A_1, \ldots, A_k)$ represents a block-diagonal matrix with $A_1, \ldots, A_k$ along the diagonal. Using \eqref{eq:em1}, we have that $S_D = \text{bdiag} (i_{\text{com}, \theta}^{-1}i_{\text{obs}, \theta}, i_{\text{com}, \psi}^{-1}i_{\text{obs}, \psi})$. If $M_{\dem}$ is the DEM map that maps $\theta_{t_j}$ to $\theta_{t_{j+1}}$, then $S_{\dem} = i_{\text{com}, \theta}^{-1}i_{\text{obs}, \theta}$ and $DM_{\text{DEM}} = I - i_{\text{com}, \theta}^{-1}i_{\text{obs}, \theta}$ because the first $P$ elements of $\theta_j^D$ equal $\theta_{t_j}$.

The speed matrices $S_{\ema}$ and $S_{\dem}$ are related using  $i_{\text{com}, \psi}$, $i_{\text{obs}, \psi}$, and $i_{\text{com}, \theta}$. If $(i_{\text{com}, \psi})_{kk}$  represents the $k$-th diagonal block of  $i_{\text{com}, \psi}$ in \eqref{eq:em2}, then  $i_{\text{com}, \bar \psi} = \sum_{k=N+1}^K (i_{\text{com}, \psi})_{kk}$ and $i_{\text{obs}, \bar \psi} =\sum_{k=N+1}^K (i_{\text{obs}, \bar \psi})_{kk}$ respectively are the complete-data and observed-data information matrices obtained using the $(1-\gamma)$-fraction of the full data ignored by the DEM subsequence generated using $M_{\dem}$.  The analytic forms of $i_{\text{com}}$ and $i_{\text{obs}}$ in \eqref{eq:em1} imply that $i_{\text{com}} = i_{\text{com}, \theta} + i_{\text{com}, \bar \psi}$ and $i_{\text{obs}} = i_{\text{obs}, \theta} + i_{\text{obs}, \bar \psi}$. The following theorem relates these information matrices to the global speeds of EM and DEM for the parameter sequence $\theta_{t_0}, \theta_{t_1}, \ldots, \theta_{t_\infty}=\hat \theta$ that are computed using the objective in \eqref{m-step-lbl} with $U_{t_j} = \{1, \ldots, N\}, j \geq 0$ in the M step.
\begin{theorem} \label{conv-rate-dem}
  Let $\{\theta_{t_j}, j \geq 0\}$ be the DEM subsequence generated using $M_{\dem}$. If Assumptions \ref{a1}--\ref{a6} hold, $\theta_{t_j}$ is in a small neighborhood around $\hat \theta$, $C_{\theta, \bar \psi} = i_{\text{com}, \theta}^{-1}i_{\text{com}, \bar \psi}$, and $O_{\bar \psi} = i_{\text{com}}^{-1}  i_{\text{obs}, \bar \psi}$, then $S_{\text{EM}} = (I + C_{\theta, \bar \psi})^{-1} S_{\text{DEM}} + O_{\bar \psi}$ and $\frac{\lambda_{\text{min}}(S_{\text{DEM}})}{1 + \lambda_{\text{max}}(C_{\theta, \bar \psi})} + \lambda_{\text{min}}(O_{\bar \psi}) \leq  \lambda_{\text{min}}(S_{\text{EM}}) \leq \frac{\lambda_{\text{min}}(S_{\text{DEM}}) }{1 + \lambda_{\text{min}} (C_{\theta, \bar \psi})} + \lambda_{\text{min}}(O_{\bar \psi})$. 
\end{theorem}
We interpret $O_{\bar \psi}$ as the (matrix) fraction of observed-data information ignored by the DEM in its fractional updates. Since $C_{\theta, \bar \psi}$ is a product of two positive semi-definite matrices, $\lambda_{\text{max}}(C_{\theta, \bar \psi}) \geq \lambda_{\text{min}}(C_{\theta, \bar \psi}) \geq 0$ and  Theorem \ref{conv-rate-dem} implies that $\lambda_{\text{min}}(S_{\text{EM}}) - \lambda_{\text{min}}(O_{\bar \psi}) \leq \lambda_{\text{min}}(S_{\text{DEM}}) $. With our interpretation of $O_{\bar \psi}$, this says that DEM cannot be slower than an EM that only uses a $\gamma$-fraction of the full data. This is true even if the DEM and EM converge to different values. 

\section{Experiments}
\label{expr}

\subsection{Setup}

We evaluate the performance of DEM in fitting linear mixed-effects models in large sample settings using the setup in \citet{Van00}. Let $p$, $q$, $m$, $n$, and $n_i$ be the number of fixed effects, number of random effects, sample size, total number of observations, and total number of observations for sample $i$ ($i=1, \ldots, m$) so that $n = \sum_{i=1}^m n_i$.  If $\yb_i \in \RR^{n_i}$ is the observation for sample $i$ for $i=1, \ldots, m$, then 
\begin{align}
  \label{eq:dat1}
  \yb_i = X_i \betab + Z_i \bb_i + \eb_i, \quad \bb_i \sim N_q(\zero, \Sigma), \, \Sigma = \tau^2 D, \quad \eb_i \sim N_{n_i}(\zero, \tau^2 I_{n_i}), 
\end{align}
where $X_i \in \RR^{n_i \times p}$ and $Z_i \in \RR^{n_i \times q}$ are known matrices of fixed and random effects covariates, respectively, $\betab \in \RR^{p}$ is the fixed effects parameter vector, $\tau^2$ is the error variance parameter, $D$ is a symmetric positive definite matrix, $\bb_i \in \RR^q$ is the random effects vector for sample $i$ that follows a $q$-dimensional Gaussian distribution with mean $\zero$ and covariance parameter $\Sigma = \tau^2 D$, and $I_{n_i}$ is $n_i$-by-$n_i$ identity matrix. \citet{Van00} developed many efficient extensions of EM-type algorithms for the estimation of $\theta = \{\betab, \Sigma, \tau^2\}$, but every extension is slow if $m$ is large due to the time consuming E step. 

We extended van Dyk's ECME algorithm, called ECME$_0$, using DEM. We randomly partitioned the $m$ samples into $K$ disjoint subsets such that observations specific to a sample were in the same subset. 
{The model in \eqref{eq:dat1} satisfies Assumptions \ref{a1}--\ref{a5} in Theorem \ref{conv-dem}; see the supplementary material for details.} DEM ran using one manager and $K$ worker processes. The E step of ECME$_0$ algorithm was split into local E steps of DEM on $K$ workers, where as the M step of DEM was performed using \eqref{m-step} on the manager. We chose three values of $\gamma = 0.3, 0.5, 0.7$ to demonstrate the trade-off between the number of iterations required to reach a local mode, faster local E steps, and the communication overhead. DEM reduced to IEM when $\gamma = 1/m$, but convergence to the local mode was too slow, so we used DEM results with $\gamma = 1/K$ as IEM results. The maximum number of iterations in any ECME$_0$, IEM, or DEM run was fixed at $10^3$, and convergence to the local mode was achieved if the change in log likelihood between two successive iterations was less than $10^{-7}$. {We implemented ECME$_0$, IEM, and DEM algorithms in R and used $\betab_0 = \zero$, $D_0 = I_q$, and $\tau_0^2 = 10$ as the starting points of these algorithms in simulated and real data analyses.} All experiments ran on a Sun Grid Engine cluster with 2.6GHz 16 core compute nodes and the $K+1$ processes in IEM and DEM algorithms  were reserved using the Rmpi package. We remark that our choice of MPI was driven by our familiarity with the Rmpi package and our DEM implementation could be significantly improved using other interfaces.

The parameter estimates, log likelihood, and run-time of ECME$_0$ algorithm served as the benchmark in all our comparisons. We compared DEM's performance with IEM, lme4, a state-of-the-art method for estimation of parameters in \eqref{eq:dat1}, and an approach following \citet{Liuetal15} that was based on meta analysis and lme4 (Meta-lme4). In Meta-lme4, $K$ mixed-effects models were fit on $K$ subsets using lme4 and the final estimate of a parameter was the average of the $K$ estimates obtained using $K$ subsets. The accuracy of every algorithm in parameter estimation was judged using errors defined as
\begin{align}
  \label{eq:met1}
  \text{err}^2_{\betab} &= p^{-1} \sum_{i=1}^p \left( \hat \beta_i - \hat \beta_i^{\text{EM}} \right)^2, \quad
  \text{err}^2_{\tau^2} = (\hat \tau^2 - \hat \tau^{2\text{EM}})^2, \quad 
  \text{err}^2_{\text{var}} = q^{-1} \sum_{i=1}^q \left( \hat \Sigma_{ii} - \hat \Sigma_{ii}^{\text{EM}} \right)^2, \nonumber \\
  \text{err}^2_{\text{cov}} &= 2q^{-1}(q-1)^{-1} \sum_{i=1}^{q-1} \sum_{j=i+1}^q \left( \hat \Sigma_{ij} - \hat \Sigma_{ij}^{\text{EM}} \right)^2, \;
\end{align}
where $\{\hat \betab^{\text{EM}}, \hat \Sigma^{\text{EM}}, \hat \tau^{2\text{EM}}\}$ and $\{\hat \betab, \hat \Sigma, \hat \tau^{2}\}$ respectively were the parameter estimates of ECME$_0$ and its competitor, including lme4, Meta-lme4, IEM, or DEM. If err$_i$ represented the error in replication $i$ of the experiment, the root mean square error (RMSE) over $R$ replications was defined as $\text{RMSE}^2_{\text{par}} = R^{-1} \sum_{i=1}^R \text{err}^2_{\text{par} \, i}$, where $\text{par} = \{\betab, \tau^2, \text{var}, \text{cov}\}$. The smaller the RMSE, the closer are the results to the benchmark ECME$_0$ algorithm. Tables comparing the RMSEs in simulated and real data analyses are in the supplementary material.  

\subsection{Simulated data analysis}

\begin{figure}[t]
  \centering
  \includegraphics[scale=0.20]{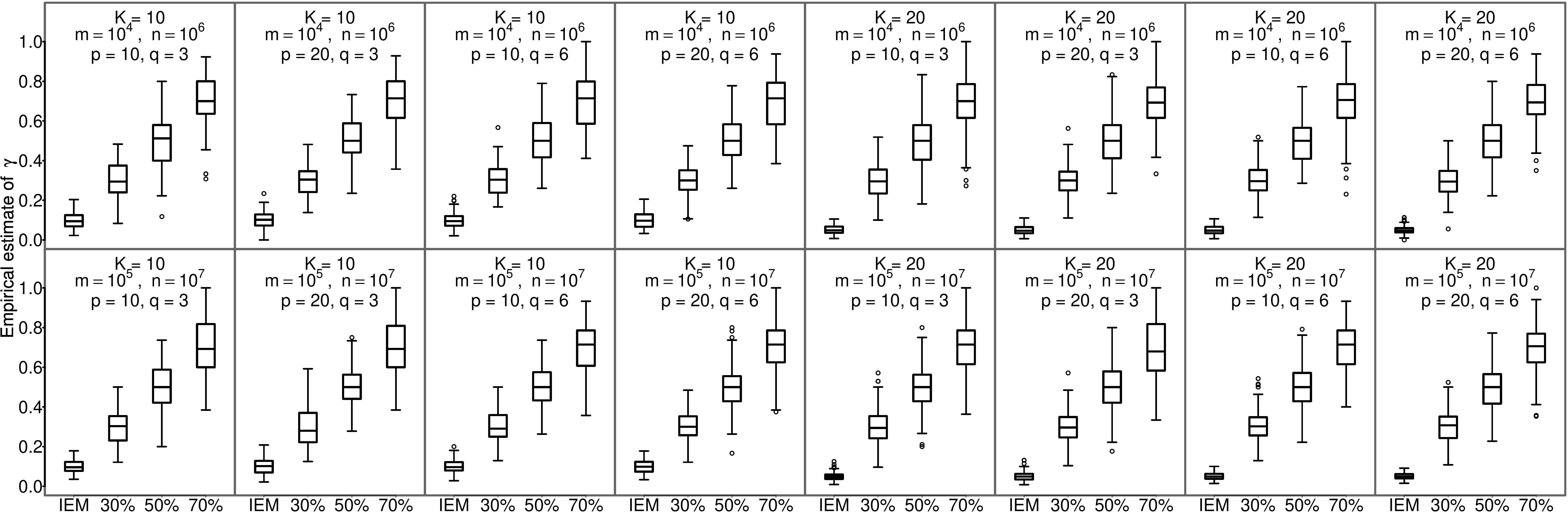}
  \caption{The empirical estimate of $\gamma$. The percentages on the x-axis represent the $\gamma$ in DEM ($\gamma = 0.3, 0.5, 0.7$). The y-axis represents the fraction of times the results communicated by the workers to the manager were accepted.}    
  \label{fig:mixef-track}
\end{figure}

We evaluated the performance of DEM on data simulated using \eqref{eq:dat1} for two  values of $(m, n)$, $p$, and $q$, respectively. We varied $(m, n) \in \{(10^4, 10^6), (10^5, 10^7)\}$, $p \in \{10, 20\}$, and  $q\in \{3, 6\}$, and randomly assigned the $n$ observations to $m$ samples. The entries of covariate matrices, $X_i$ and $Z_i$, were randomly set to $1$ or $-1$ for every $i$, $\betab$ entries were alternately fixed at $-2$ and 2, and $\tau^2$ was fixed at 1. The matrix $\Sigma = V R V^T$ if $q=3$ and $\Sigma = \text{bdiag}(V R V^T, VRV^T) $ if $q=6$, where $R$ was a $3$-by-$3$ correlation matrix with $R_{12}=-0.4$, $R_{13}=0.30$, and $R_{23}=0.001$, $V = \diag(\sqrt{1}, \sqrt{2}, \sqrt{3})$, $\diag(\ab)$ was a diagonal matrix with $\ab$ along the diagonal, and $\text{bdiag}(A_1, A_2)$ was a block-diagonal matrix with $A_1$ and $A_2$ along the diagonals. \citet{Kimetal13} showed empirically that $R$ was ideal for mixed-effects model simulations because it included negative, positive, and small to moderate strength correlations. This setup was replicated ten times for every combination of $(m, n)$, $p$, and $q$.

The simulation includes easy and hard examples for parameter estimation in \eqref{eq:dat1}. Irrespective of the  value of $p$, parameter estimation using any algorithm is efficient if $q=3$ and time-consuming if $q=6$ due to the estimation of $\Sigma$. In Meta-lme4, IEM, and DEM applications, we also present results for $K=10, 20$ to demonstrate the effect of $K$ on parameter estimates and run-time of DEM. For a fixed $\gamma$, if $K$ increases, then the local E steps on workers are faster due to smaller subset sizes but the communication overhead among workers and manager increases; however,  DEM is still faster relative to ECME$_0$ for $\gamma=0.5, 0.7$. The empirical estimates of $\gamma$ for all the workers are close to their true values across all settings (Figure \ref{fig:mixef-track}), providing an empirical confirmation of the Assumption 6 in Theorem \ref{conv-dem}. 

DEM was accurate in parameter estimation and its accuracy did not depend on the choices of $\gamma$ and $K$ across all simulation settings; see RMSEs in supplementary material. DEM outperformed its competitors, except lme4, in the estimation of  $\Sigma$ and $\tau^2$. The accuracies of Meta-lme4 and IEM were sensitive to the choice of $K$. The accuracies of lme4 and DEM were the same across all replications and $\gamma$s; however, DEM was faster than lme4 for large $m$ and was more general. We concluded that DEM's performance was closest to that of ECME$_0$ for every choice of $\gamma$, $m$, and $K$. 

\begin{figure}[t]
  \centering
  \includegraphics[scale=0.20]{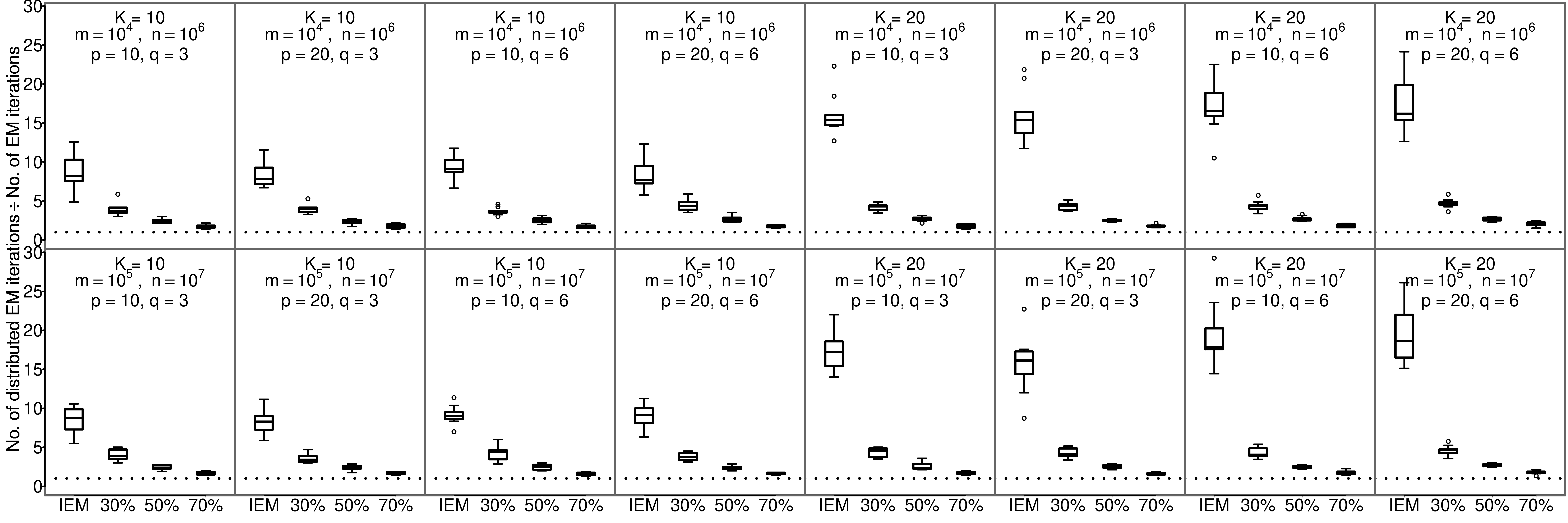}
  \caption{The number of iterations until convergence of the distributed EM relative to that of the classical EM. The percentages on the x-axis represent the $\gamma$ in DEM ($\gamma = 0.3, 0.5, 0.7$). The y-axis represents the ratio of number of iterations required by distributed EM until convergence over that of EMCE$_0$. The dotted horizontal line represents 1.}    
  \label{fig:mixef-iters}
\end{figure}

\begin{figure}[t]
  \centering
  \includegraphics[scale=0.20]{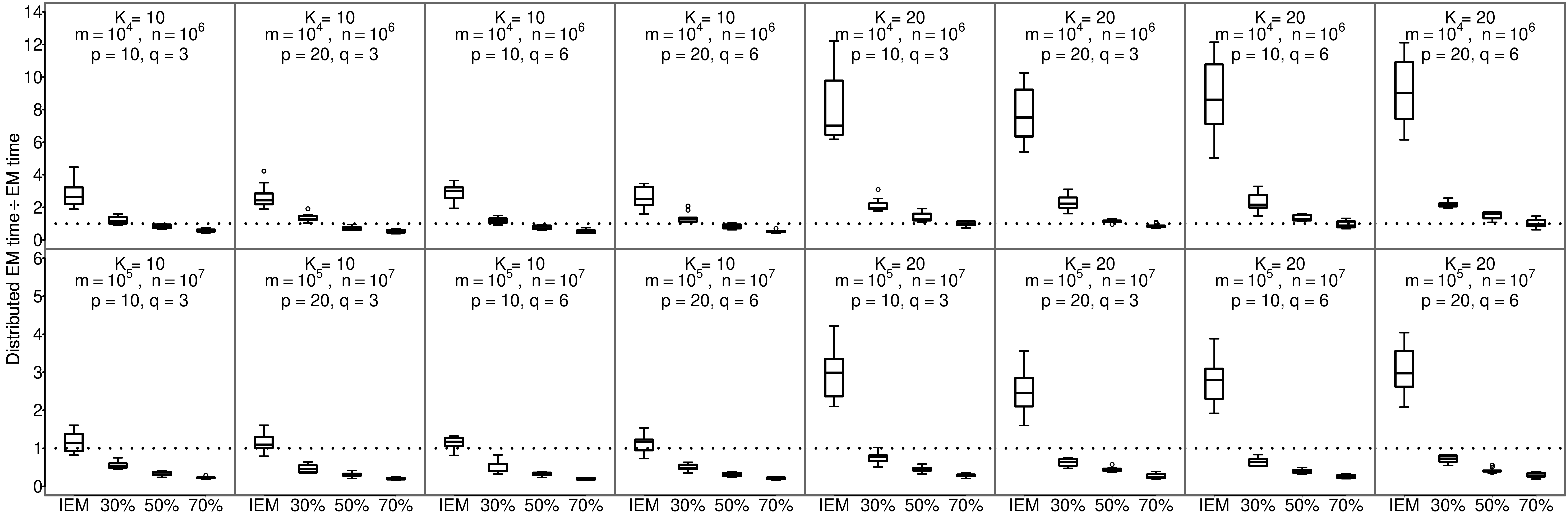}
  \caption{The time required until convergence of the distributed EM relative to that of the classical EM. The percentages on the x-axis represent the $\gamma$ in DEM ($\gamma = 0.3, 0.5, 0.7$). The y-axis represents the ratio of time required by distributed EM until convergence over that of EMCE$_0$. The dotted horizontal line represents 1. }    
  \label{fig:mixef-time}
\end{figure}

IEM and DEM were distributed generalizations of ECME$_0$, so we compared their performance in terms of log likelihood, number of iterations until convergence, and time until convergence relative to ECME$_0$. The log likehood of IEM and DEM were equal to the log likehood of ECME$_0$ across all  replications. The number of iterations required to reach a local mode by IEM or DEM were larger than that of ECME$_0$ across all simulation settings (Figure \ref{fig:mixef-iters}), providing an empirical confirmation of Theorem \ref{conv-rate-dem}; however, the increase was very small ($1.5$--$2$ times relative to ECME$_0$) for DEM with $\gamma = 0.5, 0.7$.  This implied that faster local E steps of DEM resulted in faster run-times for DEM than ECME$_0$ (Figure \ref{fig:mixef-time}). 

The differences between the run-times of DEM and IEM highlights the importance of $\gamma$. First, for a fixed $K$ and $\gamma = 0.5, 0.7$, DEM becomes faster relative to ECME$_0$ as $m$ increases due to faster local E steps and smaller time per iteration. For smaller $\gamma$s, the increased number of iterations required for convergence offsets the run-time benefits of faster E steps.  Second, for a fixed $m$ and $K$, DEM is faster as $\gamma$ increases due to the smaller number of iterations required until convergence; however, if $\gamma$ increases beyond a threshold, then the increased cost of communication offsets the run-time gains from quick convergence. In the extreme case, when $\gamma = 1$ DEM reduces to a distributed-ECME$_0$, which is slower than ECME$_0$ due to the extra communication cost. The second observation matters the most in practice because $m$ and $K$ remain fixed and $\gamma$ is chosen to balance the communication overhead and the increase in number of iterations. Due to this, IEM is slower and DEM with $\gamma = 0.7$ is faster than ECME$_0$ across all simulation settings. Since DEM with $\gamma = 0.7$ is accurate and fast, we conclude that it performs the best among all EM-type competitors.

\subsection{Real data analysis: MovieLens ratings database}
\label{real-data}

MovieLens data contain 10,000,054 ratings for 10,681 movies by 71,567 users of the online movie recommender service MovieLens (\url{http://grouplens.org}). The rating of any movie varies from 0.5 to 5 in increments of 0.5. Every observation in the database contains information about the user $i$, movie $j$, the rating $r_{ij}$ assigned by user $i$ to movie $j$, the time of rating, and genre of the movie, which could be one or more of the 19 possible categories. The MovieLens data are an example of data with repeated measures, where every user has rated at least 20 movies. If the interest lies in recommendation of movies to a user, then we can use the linear mixed-effects model in \eqref{eq:dat1} with ratings of user $i$ as the response, movie genres as the fixed and random effects covariates, and  random effects specific to the user. Agreeing with our simulations, fitting linear mixed-effects model to MovieLens data using existing tools is inefficient simply due to the large number of users and movies. 

\citet{Per17} proposed a computationally efficient approach for fitting linear mixed-effects models in large sample settings. Following \citet{Per17}, we modified the MovieLens data as follows. The $i$th response was defined as $\yb^T_i = (r_{i1}, \ldots, r_{in_i})$ $(i=1, \ldots, m)$, where $n_i$ was the number of movies rated by user $i$ and $m = 71567$. The 19 movie genres were mapped to four categories defining the \emph{genre} predictor. The Action, Adventure, Fantasy, Horror, Sci-Fi, and Thriller genres were assigned to the \emph{Action} category; the Animation and Children genres were assigned to the \emph{Children} category; the Comedy genre was assigned to the \emph{Comedy} category; and the Crime, Documentary, Drama, Film-Noir, Musical, Mystery, Romance, War, and Western genres were assigned to the \emph{Drama} category. For any movie, all genres assigned to it within a category were averaged. The \emph{movie popularity} predictor was defined to be $\mathrm{logit}\{(l + 0.5) / (n + 1.0) \}$, where $n$ was the number of ratings for the movie in 30 most recent observations for the movie and $l$ was the number of users who rated the movie higher than $3$. The \emph{previous} predictor was defined to be 1 if the user rated the previous movie to be larger than $3$ and 0 otherwise. The $X_i$ and $Z_i$ matrices in \eqref{eq:dat1} were defined based on the \emph{genre},  \emph{movie popularity}, and \emph{previous} predictors and each had six columns. 

DEM led to an easy extension of ECME$_0$ algorithm for the analysis of Perry's data set. We randomly divided the users in Perry's data into 10 sets of training data. All ratings specific to a user were contained in the same training data. We ran 10 replications of our experiments for the 10 training datasets. We randomly divided the samples in training data into 20 disjoint subsets, reserved $21$ processes on a cluster, one for the manager and the other 20 for the workers, and stored the 20 data subsets separately on the 20 worker processes. DEM matched the accuracy of ECME$_0$ in parameter estimation for every $\gamma$. The same was also true for IEM and lme4. On the other hand, Meta-lme4 was slightly inaccurate compared to its competitors. The log likelihoods of DEM for all $\gamma$s and IEM were equal to that of ECME$_0$ across all replications; see the tables in supplementary material. 

\begin{figure}[t]
  \centering
  \subfloat[]{
    \includegraphics[scale=0.20]{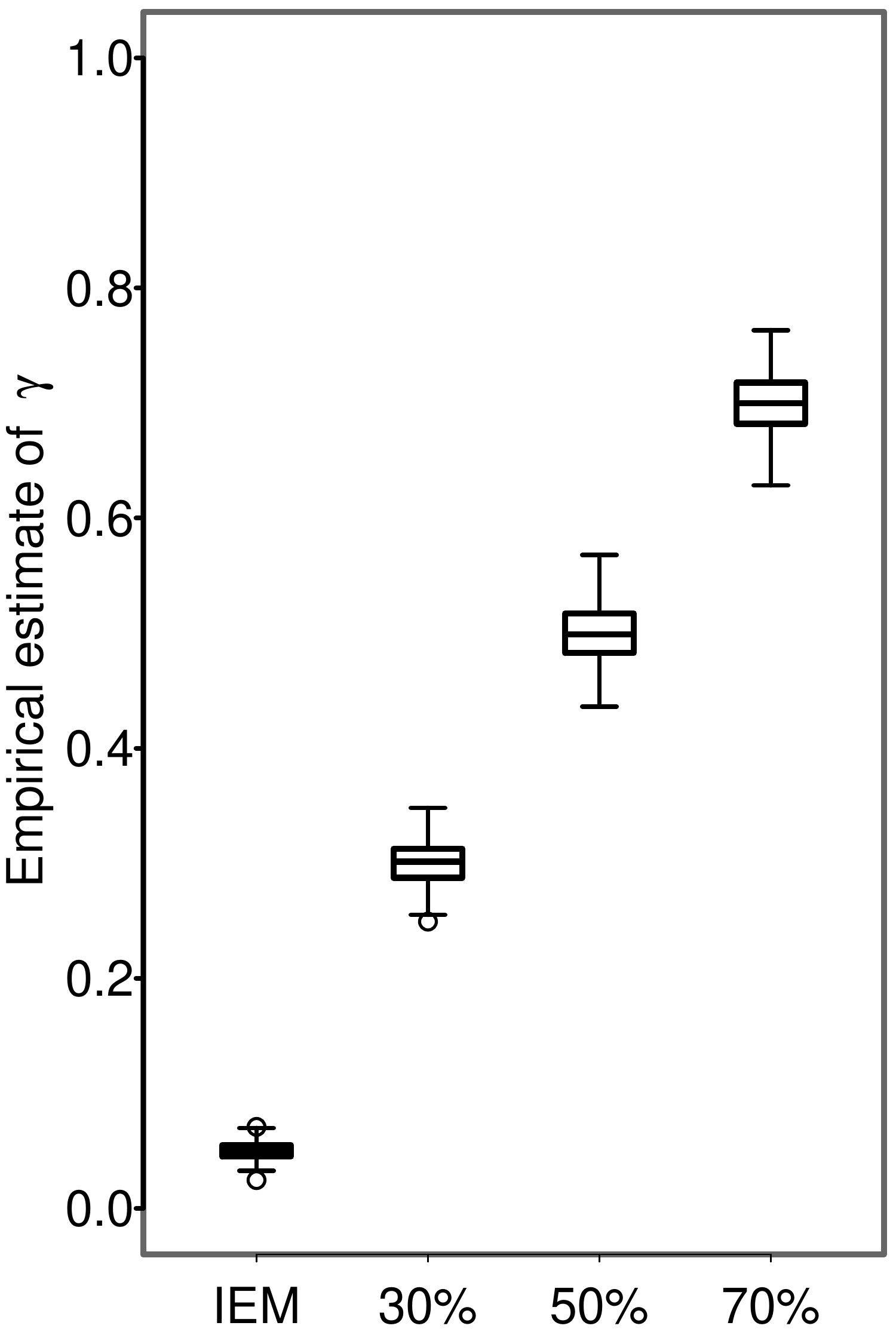}
    \label{fig:track}}\hspace{20pt}  
  \subfloat[]{
    \includegraphics[scale=0.20]{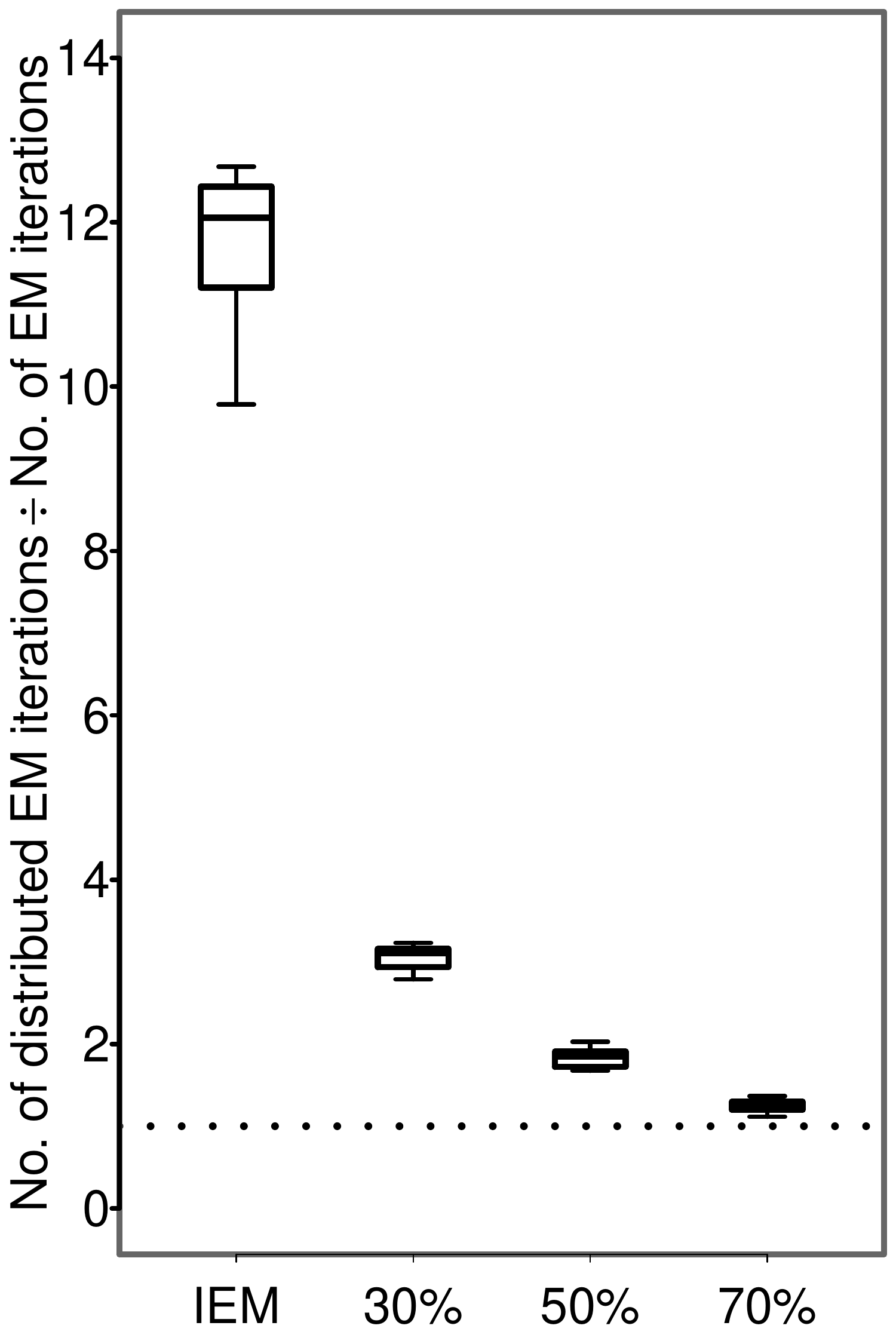}
    \label{fig:iters}} \hspace{20pt}
  \subfloat[]{
    \includegraphics[scale=0.20]{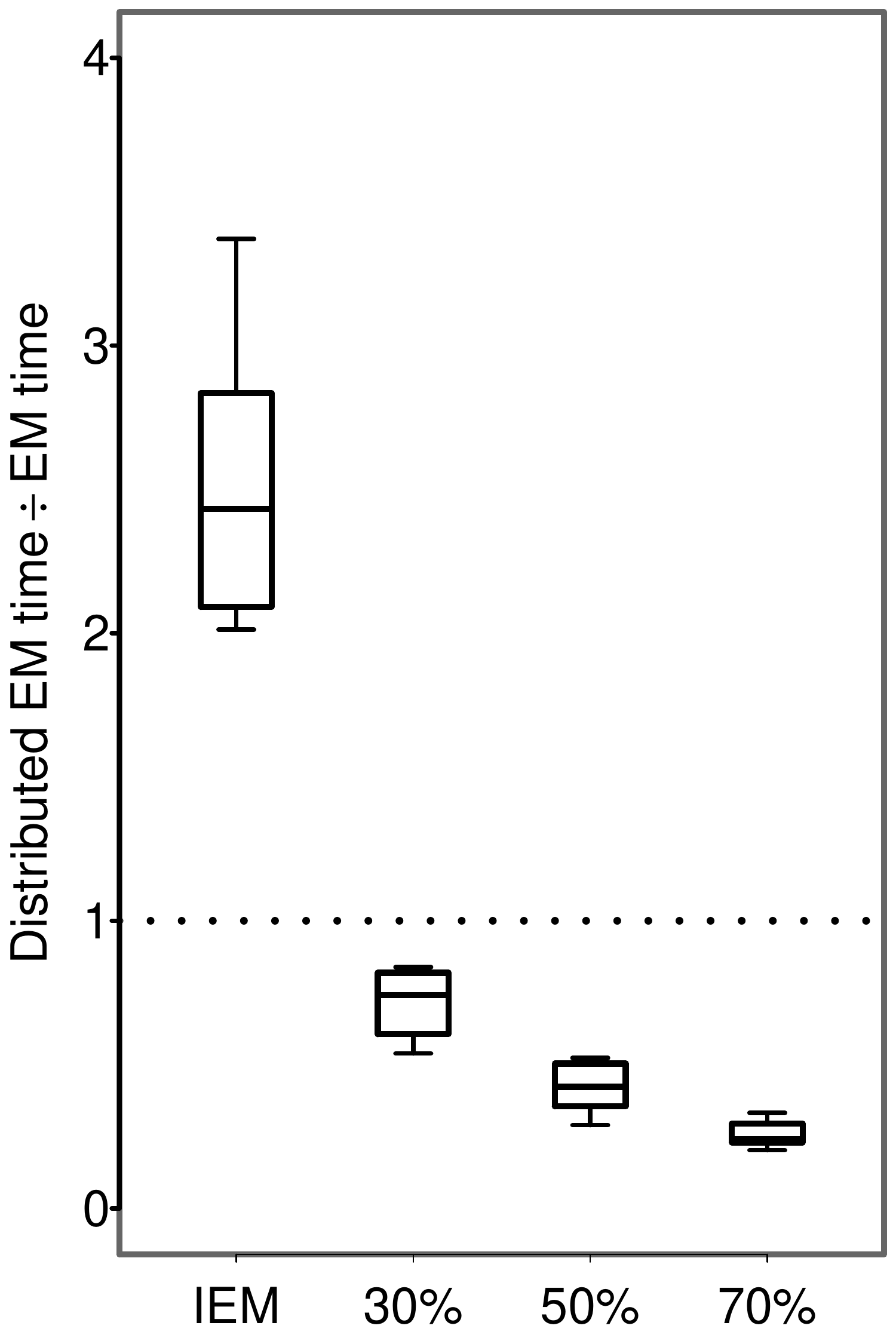}
    \label{fig:time}}
  \caption{Comparison of IEM and DEM ($\gamma = 0.3, 0.5, 0.7$) performance relative to ECME$_0$. (a) The empirical estimate of $\gamma$. (b) The ratio of the number of iterations until convergence required by IEM or DEM over that of EMCE$_0$. (c) The ratio of the IEM or DEM run-time over EMCE$_0$ run-time. The dotted horizontal line represents 1.}    
  \label{fig:dem-em}
\end{figure}

DEM is faster than ECME$_0$ and lme4 across all simulation replications and for every $\gamma$. The computational burden is different for every worker because the number of movies rated by users vary a lot; therefore, the workers with minimal burden return their $Q_i$s to the manager more often than other workers. Due to this, the variability in the empirical estimates of $\gamma$ increases with $\gamma$ (Figure \ref{fig:track}). As $\gamma$ increases, the time and the number of iterations required until convergence decrease (Figures \ref{fig:iters} and \ref{fig:time}). All these results agree closely with our simulation results; therefore, we conclude that DEM with $\gamma=0.7$ achieves the best balance of efficiency and accuracy when compared to ECME$_0$ results. 

\section{Discussion}

\begin{figure}[t]
  \centering
    \includegraphics[scale=0.20]{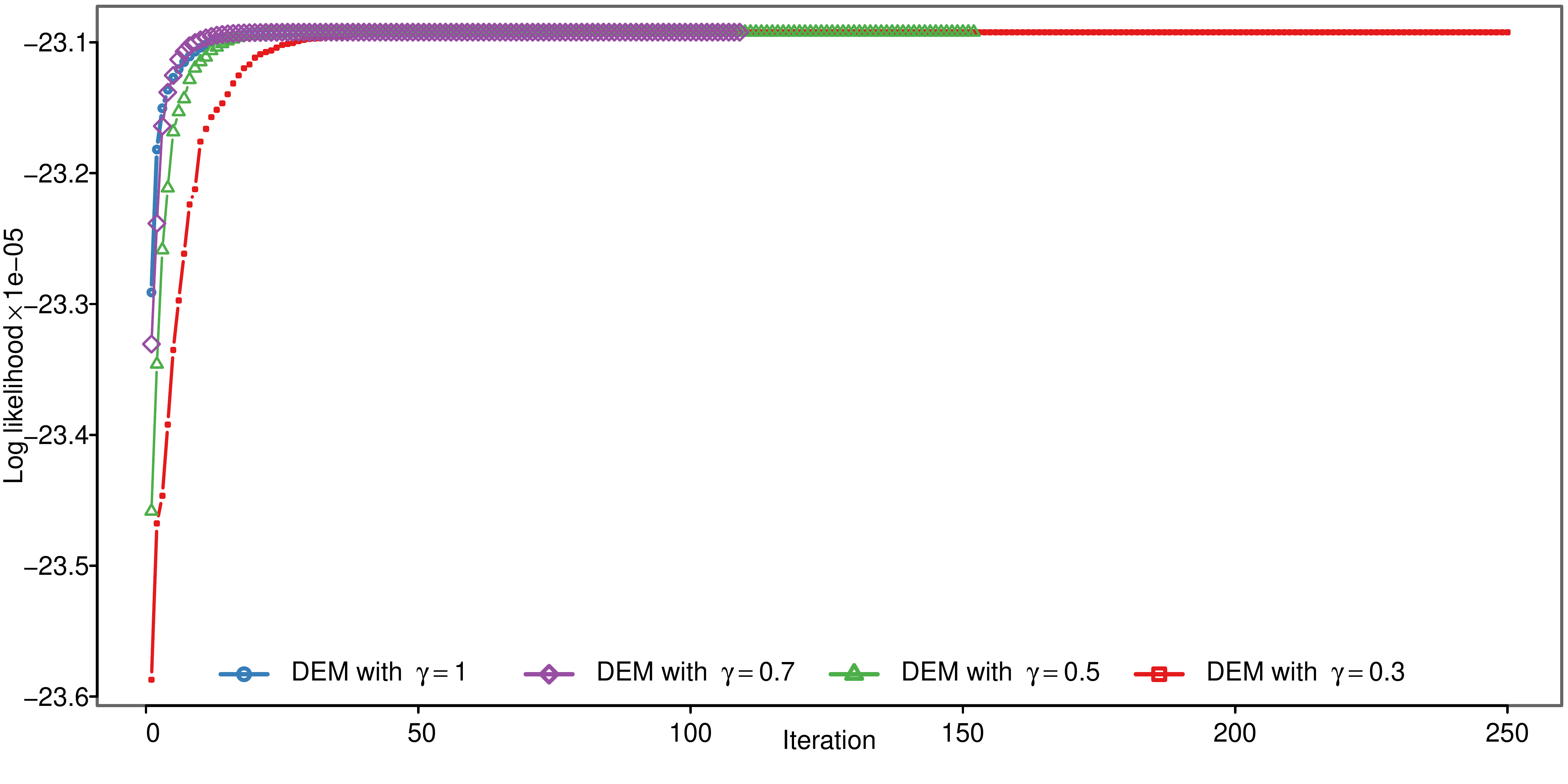}  
    \caption{The log likelihood values across different iterations of DEM with $\gamma = 0.3, 0.5, 0.7, 1$ in a replication of the MovieLens data analysis. The DEMs with $\gamma =0.3$ and $\gamma = 1$ respectively required the largest and smallest number of iterations before convergence.}    
  \label{fig:dem-log}
\end{figure}

The DEM algorithm is tuned for massive data applications using distributed computing. 
In cases where the computer cluster is small and the data cannot be loaded into memory, the DEM algorithm can be 
extended to load data subsets infrequently for improved efficiency. While we have focused on DEM applications for maximizing the log likelihood, it is also applicable for EM-type estimation in Bayesian modeling and approximate Bayesian inference. Currently we ignore the results of ($1-\gamma$)-fraction of workers which did not return their results to the manager before the M step, which wastes computational resources. A DEM extension that uses the results of local computations on workers in future DEM updates and balances computational load on the workers is an idea worth pursuing.

{The main novelty of DEM is fractional updates, but this also implies that DEM is not a GEM. Our simulations and real data analyses, however, show that the monotone ascent of the $\Lcal(\theta_t)$ sequence is rarely violated. Most violations happen when $t$ is small, and they are more common in IEM than DEM. Specifically, the $\Lcal(\theta_t)$ sequence retains its monotonic ascent in DEM with $\gamma = 0.5$ or $0.7$ across every replication of simulated and real data analysis (Figure \ref{fig:dem-log}), where it decreases multiple times in the IEM. Empirically if $\gamma$ exceeds a threshold, which in our case is 0.5, then DEM behaves as a GEM with high probability.}

We are exploring extensions of the DEM algorithm that choose $\gamma$ depending on the model complexity and sample size. Our empirical results show that the parameter estimates are robust to the choice of $\gamma$; however, if $\gamma$  is very small, then the number of iterations until convergence can be very large and there is no benefit of using DEM over its non-distributed version, such as the high run-times of IEM in real data analysis. It can be interesting to use dynamic values for $\gamma$. At earlier iterations, EM typically moves fast toward the target solution. The use of a small $\gamma$ value is expected to be sufficient. In the final iterations, large values may help to obtain more accurate estimates. This suggests that the stability of change, for example, in log likelihood for convergence monitoring can be used for dynamically specifying $\gamma$ values. 

More research is required for generalizing DEM to dependent data.  The DEM algorithm is applicable to any type of independent data, but a DEM generalization based on the divide-and-conquer technique becomes challenging when the data are dependent. Based on the development in online EMs for hidden Markov models \citep{Cap11}, solutions may  depend on specific model structures. For example, even in traditional approaches to maximum likelihood estimation for time series, the information associated with the marginal distribution for the first few data points is ignored for simplicity. But in the divide-and-conquer setting, subsets can be created with certain overlaps. We leave this for future work.

\section*{Acknowledgments}
This research was supported in part through computational resources provided by The University of Iowa, Iowa City, Iowa. Chuanhai Liu's work was partially supported by the National Science Foundation grant DMS-1316922. The code used in the experiments is available at \url{https://github.com/blayes/DEM}. 

\bibliographystyle{agsm}
\bibliography{parallelem}

\clearpage

\renewcommand\thesection{\arabic{section}}
\renewcommand\thesubsection{\thesection.\arabic{subsection}}
\renewcommand\thesubsubsection{\thesubsection.\arabic{subsubsection}}

\setcounter{section}{0}

\begin{center}
\textbf{\Large Supplementary Material for An Asynchronous Distributed Expectation Maximization Algorithm For Massive Data: The DEM Algorithm}
\end{center}

\section{Proof of Theorems in Sections 4.2 and 4.3}
\label{sec:intro}

Our theoretical setup has the following assumptions:
\begin{enumerate}    
\item  \label{a1} $\Theta$ is a subset in the $P$-dimensional Euclidean space $\RR^P$.
\item \label{a2} The set $\Pi_{\theta_0} \otimes \Theta_{\theta_0} = \{(\widetilde p, \theta) \in \Pi \otimes \Theta: \, F(\widetilde p, \theta) \ge F(\tilde p_0, \theta_0) \}$ is compact for any starting point of the ($\tilde p_t$, $\theta_t$) sequence, denoted as $(\tilde p_0, \theta_0)$, that satisfies $ \Lcal(\theta_0) > -\infty$ and $\tilde p_0 = \prod_{k=1}^K h(Y_k \mid Z_k, \theta_0)$.   
\item  \label{a3} $F(\tilde p, \theta)$ is continuous in $\Pi \otimes \Theta$ and differentiable in the interior of $\Pi \otimes \Theta$.
\item  \label{a4} $\Pi_{\theta_0} \otimes \Theta_{\theta_0}$ is in the interior of $\Pi \otimes \Theta$ for any $\theta_0 \in \Theta$.  
\item  \label{a5} The first order differential ${\partial Q(\theta \mid \theta_{t}, \theta_{t_{(N+1)}}, \ldots, \theta_{t_{K}})} / {\partial \theta}$ is continuous in $(\theta, \theta_t, \theta_{t_{(N+1)}}, \ldots, \theta_{t_{K}})$.    
\item  \label{a6} Worker $k$ returns $Q_k$ to the manager infinitely often for $t=0, \ldots, \infty$ and $k = 1, \ldots, K$.
\end{enumerate}

\subsection{Proof of Theorem 4.1}

  The proof uses arguments similar to Theorems 1 and 2 of \citet{NeaHin98}. First, the E step of DEM at the  $(t+1)$-th iteration updates $\widetilde p_{t,k} = h(Y_k \mid Z_k, \theta_{t_k})$ to $\widetilde p_{(t+1), k} = h(Y_k \mid Z_k, \theta_t)$ for worker $k$ if $k \in U_{(t+1)}$; otherwise, $\widetilde p_{(t+1), k} = h(Y_k | Z_k, \theta_{t_k})$. Define $\widetilde p_{(t+1)} = \prod_{k_1 \in  U_{t+1}} \widetilde p_{(t+1), k_1} \prod_{k_2 \in U_{t+1}^c} \widetilde p_{(t+1), k_2}$. Theorem 1 in \citet{NeaHin98} implies that $F(\widetilde p_{t}, \theta_{t}) \leq F(\widetilde p_{(t+1)}, \theta_{t})$ for a given $\theta_{t}$. Second, the M step of DEM at the  $(t+1)$-th iteration updates $\theta_t$ to $\theta_{(t+1)}$ and increases $F$ from $F(\widetilde p_{(t+1)}, \theta_t)$ to  $F(\widetilde p_{(t+1)}, \theta_{(t+1)})$ for fixed $\widetilde p_{(t+1)}$. At the end of $(t+1)$-th iteration of DEM, $F(\widetilde p_{t}, \theta_t) \leq F(\widetilde p_{(t+1)}, \theta_{t}) \leq F(\widetilde p_{(t+1)}, \theta_{(t+1)})$, where the first and last equality follow from Theorem 1 in \citet{NeaHin98}. Because $t$ is a generic iteration,  DEM maintains the monotone ascent of $F(\widetilde p, \theta)$ at every iteration and $\{F(\widetilde p_t, \theta_t), t \geq 0\}$ sequence converges because $F(\widetilde p, \theta)$ is upper bounded by our assumption. Theorem 2 in \citet{NeaHin98} implies that if $(\hat {\widetilde p}, \hat \theta)$ is a fixed point of $F(\widetilde p_t, \theta_t)$ sequence, then $\hat \Lcal = \Lcal(\hat \theta)$ is a fixed point of $\Lcal(\theta_t)$ sequence. This implies that there exists a monotone subsequence of $\Lcal(\theta_t)$ converging to $\hat \Lcal$. 

\subsection{Proof of Theorem 4.2}


  To prove this theorem, we require the definition of a closed map. A point-to-set mapping $A$ is closed on a set $X$ if $x_k \rightarrow x$, $x_k \in X$, and $y_k \rightarrow y$, $y_k \in A(x_k)$, then $y \in A(x)$ for every $x \in X$; see \citet[pp 203]{LueYe08} for details. If $A$ is continuous, then it is closed. 

  The proof is based on Theorems 1, 2, 4, and 5 in \citet{Wu83}. Our assumptions imply that the point-to-set map $(\tilde p_t, \theta_t) \mapsto  (\tilde p_{t+1}, \theta_{t+1})$ is continuous, thus closed, on $\Pi \otimes \Theta \cap (\Scal \cup \Mcal)^c$; see Theorem 2 in \citet{Wu83}. Theorem 2 in \citet{NeaHin98} implies that $F(\widetilde p_{t}, \theta_t) \leq F(\widetilde p_{t+1}, \theta_{t+1})$ for every $(\widetilde p_{t}, \theta_t) \in \Pi \otimes \Theta \cap (\Scal \cup \Mcal)^c$, so $F$ is our ascent function. The global convergence theorem in \cite{Wu83} implies that all limit points of $(\widetilde p_{t}, \theta_t) $ sequence lie in $ \Scal \cup \Mcal$ and  $F(\widetilde p_{t}, \theta_t)$ converges monotonically to $\hat F = F(\hat {\widetilde p}, \hat {\theta})$ for some $(\hat {\widetilde p}, \hat {\theta}) \in \Scal \cup \Mcal$.

  If $\Scal (\hat F)$ (respectively $\Mcal (\hat F)$) $ = \{(\hat {\widetilde p}, \hat {\theta})\}$ , then there cannot be two different stationary points (respectively local maxima) with the same $\hat F$. This implies that $(\widetilde p_{t}, \theta_t) \rightarrow (\hat {\widetilde p}, \hat {\theta})$ and $\theta_t \rightarrow \hat \theta$ using coordinate-wise convergence. The first part of the theorem is proved.

  Assumption \ref{a2} implies that $(\widetilde p_{t}, \theta_t)$ is a bounded sequence, so Theorem 5 in \citet{Wu83} implies that the set of limit points of the sequence $(\widetilde p_{t}, \theta_t)$ with $\|(\tilde p_{t+1}, \theta_{t+1}) - (\tilde p_t, \theta_t)\|_{\Pi \otimes \Theta} \rightarrow 0$ as $t \rightarrow \infty$ is connected and compact. Since $\Scal(\hat F)$ and $\Mcal(\hat F)$ are discrete, the only connected and compact components of the stationary points (respectively local maxima) are singletons. All the limit points of  $(\widetilde p_{t}, \theta_t)$ are in $\Scal(\hat F) \cup \Mcal(\hat F)$, so $(\widetilde p_{t}, \theta_t) \rightarrow (\hat {\widetilde p}, \hat {\theta})$ and the second part of the theorem is also proved.   

\subsection{Proof of Theorem 4.3}

Recall that
\begin{align*}
  i_{\text{com}, \bar \psi} &= \sum_{i=N+1}^K (i_{\text{com}, \psi})_{kk}, \quad i_{\text{obs},  \bar \psi} =\sum_{k=N+1}^K (i_{\text{obs}, \psi})_{kk}, \quad 
  i_{\text{com}} = i_{\text{com}, \theta} + i_{\text{com},  \bar \psi}, \quad   i_{\text{obs}} = i_{\text{obs}, \theta} + i_{\text{obs},  \bar \psi}.
\end{align*}
Define $C_{\theta, \bar \psi} = i_{\text{com}, \theta}^{-1}i_{\text{com}, \bar \psi}$ and $O_{\bar \psi} = i_{\text{com}}^{-1}  i_{\text{obs}, \bar \psi}$ and substitute them in
\begin{align*}
  S_{\ema} =  i_{\text{com}}^{-1} i_{\text{obs}},\; i_{\text{obs}} = - \tfrac{\partial^2 \log g(Z_{1:K} \mid \theta)} {\partial \theta \cdot \partial \theta^T} \big|_{\theta = \hat \theta^E}, \; i_{\text{com}} = - \EE_Y \left\{ \tfrac{\partial^2 \log f(Y_{1:K}, Z_{1:K} \mid \theta)} {\partial \theta \cdot \partial \theta^T} \mid Z_{1:K}, \theta \right\}  \big|_{\theta = \hat \theta^E}, 
\end{align*}
where $i_{\text{obs}}$ and $i_{\text{com}}$ are the observed-data and complete-data information matrices, to obtain that
\begin{align*}
  S_{\text{EM}} &= i^{-1}_{\text{com}} i_{\text{obs}} 
                  = (I + i_{\text{com}, \theta}^{-1}i_{\text{com}, \bar \psi})^{-1} i_{\text{com}, \theta}^{-1}  i_{\text{obs}, \theta} + O_{\bar \psi} = (I + C_{\theta, \bar \psi})^{-1} S_{\text{DEM}} + O_{\bar \psi}.
\end{align*}
Simplifying the equality in the above display yields
\begin{align}
  \label{eq:ac1}
    \lambda_{\text{min}}(S_{\text{EM}}) &\overset{(i)}{\leq} \lambda_{\text{max}}\{(I + C_{\theta, \bar \psi})^{-1}S_{\text{DEM}}\} + \lambda_{\text{min}}(O_{\bar \psi})  \overset{(ii)}{\leq}  
                                          \lambda_{\text{max}}\{(I + C_{\theta, \bar \psi})^{-1}\} \lambda_{\text{min}}\{S_{\text{DEM}}\} + \lambda_{\text{min}}(O_{\bar \psi}) \nonumber \\
                                        & = 
                                          \{1 + \lambda_{\text{min}}(C_{\theta, \bar \psi})\}^{-1} \lambda_{\text{min}}\{S_{\text{DEM}}\} + \lambda_{\text{min}}(O_{\bar \psi}) \\
\lambda_{\text{min}}(S_{\text{EM}}) &\overset{(iii)}{\geq}    \lambda_{\text{min}}\{(I + C_{\theta, \bar \psi})^{-1}S_{\text{DEM}}\} + \lambda_{\text{min}}(O_{\bar \psi}) \overset{(iv)}{\geq} 
\lambda_{\text{min}}\{(I + C_{\theta, \bar \psi})^{-1}\} \lambda_{\text{min}}(S_{\text{DEM}}) + \lambda_{\text{min}}(O_{\bar \psi}) \nonumber \\
&= \{1 +\lambda_{\text{max}}(C_{\theta, \bar \psi}) \}^{-1} \lambda_{\text{min}}(S_{\text{DEM}}) + \lambda_{\text{min}}(O_{\bar \psi}) ,
\end{align}
where inequalities $(i)$, $(ii)$, $(iii)$, and $(iv)$ follow from Problem III.6.5 in \citet{Bha97};
therefore,
\begin{align*}
\frac{\lambda_{\text{min}}(S_{\text{DEM}})}{1 + \lambda_{\text{max}}(C_{\theta, \bar \psi})} + \lambda_{\text{min}}(O_{\bar \psi}) \leq  \lambda_{\text{min}}(S_{\text{EM}}) \leq \frac{\lambda_{\text{min}}(S_{\text{DEM}}) }{1 + \lambda_{\text{min}} (C_{\theta, \bar \psi})} + \lambda_{\text{min}}(O_{\bar \psi}).
\end{align*}  

\section{Additional experimental results from Section 5}

Recall the linear mixed effects model used for experiments. Let $p$, $q$, $m$, $n$, and $n_i$ be the number of fixed effects, number of random effects, sample size, total number of observations, and total number of observations for sample $i$ ($i=1, \ldots, m$) so that $n = \sum_{i=1}^m n_i$.  If $\yb_i \in \RR^{n_i}$ is the observation for sample $i$ for $i=1, \ldots, m$, then 
\begin{align}
  \label{eq:dat1}
  \yb_i = X_i \betab + Z_i \bb_i + \eb_i, \quad \bb_i \sim N_q(\zero, \Sigma), \, \Sigma = \tau^2 D, \quad \eb_i \sim N_{n_i}(\zero, \tau^2 I_{n_i}), 
\end{align}
where $X_i \in \RR^{n_i \times p}$ and $Z_i \in \RR^{n_i \times q}$ are known matrices of fixed and random effects covariates, respectively, $\betab \in \RR^{p}$ is the fixed effects parameter vector, $\tau^2$ is the error variance parameter, $D$ is a symmetric positive definite matrix, $\bb_i \in \RR^q$ is the random effects vector for sample $i$ that follows a $q$-dimensional Gaussian distribution with mean $\zero$ and covariance parameter $\Sigma = \tau^2 D$, and $I_{n_i}$ is $n_i$-by-$n_i$ identity matrix. The parameter vector is $\theta = \{\betab, \Sigma, \tau^2\}$. 

{The linear mixed-effects model in \eqref{eq:dat1} satisfies Assumptions \ref{a1}--\ref{a5} in Theorem 4.2. Let $LL^T$ be the Cholesky decomposition of $\Sigma$, where $L$ is lower triangular, and $\text{vech}(L)$ be the lower triangular part of $L$ arranged in a $q(q+1)/2$-dimensional vector. Our parameter vector can be also defined as $\theta = \{\betab, \text{vech}(L), \tau^2\}$ and we assume that the parameter space $\Theta$ is a compact subset of the $(p + q(q+1)/2 + 1)$-dimensional Euclidean space. This verifies Assumption \ref{a1}. In our simulation and real data analysis, we fix $\betab_0 = \zero$, $L_0 = I_q$, and $\tau_0^2 = 10$ as the starting point of DEM iterations. The conditional distribution of missing data $\bb_i$ in \eqref{eq:dat1} is also Gaussian with mean $\hat \bb_i$ and covariance matrix $\hat C_i$ ($i=1, \ldots, m$); see Equation 3.6 in \citet{Van00} for the analytic forms of $\hat \bb_i$ and $\hat C_i$. Define $\Pi$ in Assumption \ref{a2} to be a compact set of continuous distributions with  density $\tilde p$, finite  KL$\{\tilde p, N(\hat \bb_i, \hat C_i)\}$ for every $i$, and finite $\int \log \{\tilde p(y) \} \tilde p(y) dy$. For any such $\theta_0$,  $\Pi_{\theta_0} \otimes \Theta_{\theta_0}$ is a compact subset of $\Pi \otimes \Theta$, which verifies Assumption \ref{a4}. Assumption \ref{a2} is true because the likelihood function is finite at $\theta_0 = \{\betab_0, \text{vech}(L_0), \tau^2_0\}$. The likelihood for $\theta$ in \eqref{eq:dat1} is based on a Gaussian density and is differentiable in the interior of $\Theta$, which verifies Assumption \ref{a3}. Equation 3.4 in \citet{Van00} shows that $Q_k(\theta \mid \theta_{t_k})$ is differentiable for every $k$. The $Q$-function in DEM is the sum of  $Q_1(\theta \mid \theta_{t_1}), \ldots, Q_K(\theta \mid \theta_{t_K})$, so it is also differentiable, which verifies Assumption \ref{a5}. Our implementation ensures that $Q_k(\theta \mid \theta_{t_k})$ is returned to the manager for every $k$ before convergence is declared, satisfying Assumption \ref{a6}.}

The accuracy of every algorithm in parameter estimation was judged using errors defined as
\begin{align}
  \label{eq:met1}
  \text{err}^2_{\betab} &= p^{-1} \sum_{i=1}^p \left( \hat \beta_i - \hat \beta_i^{\text{EM}} \right)^2, \quad
  \text{err}^2_{\tau^2} = (\hat \tau^2 - \hat \tau^{2\text{EM}})^2, \quad 
  \text{err}^2_{\text{var}} = q^{-1} \sum_{i=1}^q \left( \hat \Sigma_{ii} - \hat \Sigma_{ii}^{\text{EM}} \right)^2, \nonumber \\
  \text{err}^2_{\text{cov}} &= 2q^{-1}(q-1)^{-1} \sum_{i=1}^{q-1} \sum_{j=i+1}^q \left( \hat \Sigma_{ij} - \hat \Sigma_{ij}^{\text{EM}} \right)^2, \;
\end{align}
where $\{\hat \betab^{\text{EM}}, \hat \Sigma^{\text{EM}}, \hat \tau^{2\text{EM}}\}$ and $\{\hat \betab, \hat \Sigma, \hat \tau^{2}\}$ respectively were the parameter estimates of ECME$_0$ and its competitor, including lme4, Meta-lme4, IEM, or DEM. If err$_i$ represented the error in replication $i$
of the experiment, the root mean square error (RMSE) over $R$ replications was defined as
\begin{align}
  \label{eq:met2}
  \text{RMSE}^2_{\betab} = R^{-1} \sum_{i=1}^R \text{err}^2_{\betab \, i}, \quad
  \text{RMSE}^2_{\tau^2} = R^{-1} \sum_{i=1}^R \text{err}^2_{\tau^2\,  i}, \nonumber\\ 
  \text{RMSE}^2_{\text{var}} = R^{-1} \sum_{i=1}^R \text{err}^2_{\text{var}\,  i}, \quad 
  \text{RMSE}^2_{\text{cov}} = R^{-1} \sum_{i=1}^R \text{err}^2_{\text{cov}\,  i}. 
\end{align}
The smaller the RMSE, the closer are the results to the benchmark ECME$_0$ algorithm.

\begin{table}[ht]
\caption{Root mean square error \eqref{eq:met2}  in estimation of fixed effects ($\betab$) averaged across simulation replications. The maximum Monte Carlo error is of the order $10^{-4}$}
\label{tab:fix}
  \centering
{\scriptsize
  \begin{tabular}{|r|c|c|c|c|c|c|c|c|}
    \hline
     & \multicolumn{8}{c|}{$K = 10$} \\
    \hline
     & \multicolumn{4}{c|}{$m = 10^4, n = 10^6$} & \multicolumn{4}{c|} {$m = 10^5, n = 10^7$} \\
    \hline
     & \multicolumn{2}{c|}{$q = 3$} & \multicolumn{2}{c|} {$q = 6$} & \multicolumn{2}{c|}{$q = 3$} & \multicolumn{2}{c|} {$q = 6$} \\
    \hline
     & $p=10$ & $p=20$ & $p=10$ & $p=20$ & $p=10$ & $p=20$ & $p=10$ & $p=20$ \\
    \hline
    lme4 & 0.0000 & 0.0000 & 0.0000 & 0.0000 & 0.0000 & 0.0000 & 0.0000 & 0.0000 \\ 
    Meta-lme4 & 0.0000 & 0.0000 & 0.0000 & 0.0000 & 0.0000 & 0.0000 & 0.0000 & 0.0000\\ 
    IEM & 0.0000 & 0.0000 & 0.0000 & 0.0000 & 0.0000 & 0.0000 & 0.0000 & 0.0000\\ 
    DEM ($\gamma = 0.3$) & 0.0000 & 0.0000 & 0.0000 & 0.0000 & 0.0000 & 0.0000 & 0.0000 & 0.0000\\ 
    DEM ($\gamma = 0.5$) & 0.0000 & 0.0000 & 0.0000 & 0.0000 & 0.0000 & 0.0000 & 0.0000 & 0.0000\\ 
    DEM ($\gamma = 0.7$) & 0.0000 & 0.0000 & 0.0000 & 0.0000 & 0.0000 & 0.0000 & 0.0000 & 0.0000\\
    \hline
    & \multicolumn{8}{c|}{$K = 20$} \\
    \hline
    & \multicolumn{4}{c|}{$m = 10^4, n = 10^6$} & \multicolumn{4}{c|} {$m = 10^5, n = 10^7$} \\
    \hline
    & \multicolumn{2}{c|}{$q = 3$} & \multicolumn{2}{c|} {$q = 6$} & \multicolumn{2}{c|}{$q = 3$} & \multicolumn{2}{c|} {$q = 6$} \\
    \hline
    & $p=10$ & $p=20$ & $p=10$ & $p=20$ & $p=10$ & $p=20$ & $p=10$ & $p=20$ \\
    \hline
    lme4 & 0.0000 & 0.0000 & 0.0000 & 0.0000 & 0.0000 & 0.0000 & 0.0000 & 0.0000 \\ 
    Meta-lme4 & 0.0001 & 0.0000 & 0.0001 & 0.0000 & 0.0000 & 0.0000 & 0.0000 & 0.0000 \\ 
    IEM & 0.0000 & 0.0000 & 0.0000 & 0.0000 & 0.0000 & 0.0000 & 0.0000 & 0.0000 \\ 
    DEM ($\gamma = 0.3$) & 0.0000 & 0.0000 & 0.0000 & 0.0000 & 0.0000 & 0.0000 & 0.0000 & 0.0000 \\ 
    DEM ($\gamma = 0.5$) & 0.0000 & 0.0000 & 0.0000 & 0.0000 & 0.0000 & 0.0000 & 0.0000 & 0.0000 \\ 
    DEM ($\gamma = 0.7$) & 0.0000 & 0.0000 & 0.0000 & 0.0000 & 0.0000 & 0.0000 & 0.0000 & 0.0000 \\ 
    \hline
  \end{tabular}
}%
\end{table}

\begin{table}[ht]
  \caption{Root mean square error \eqref{eq:met2} in the estimation of $\tau^2$ and $\Sigma$ averaged across simulation replications. The maximum Monte Carlo errors are of the order $10^{-4}$, $10^{-2}$, and $10^{-3}$ for the error variances, variances of the random effects, and covariances of random effects}  
  \label{tab:cov}
  \centering
  {\scriptsize    
    \begin{tabular}{|r|c|c|c|c|c|c|c|c|}
      \hline
      & \multicolumn{8}{c|}{\textbf{Error variance} ($\tau^2$)} \\
      \hline
      & \multicolumn{8}{c|}{$K = 10$} \\
      \hline
      & \multicolumn{4}{c|}{$m = 10^4, n = 10^6$} & \multicolumn{4}{c|} {$m = 10^5, n = 10^7$} \\
      \hline
      & \multicolumn{2}{c|}{$q = 3$} & \multicolumn{2}{c|} {$q = 6$} & \multicolumn{2}{c|}{$q = 3$} & \multicolumn{2}{c|} {$q = 6$} \\
      \hline
      & $p=10$ & $p=20$ & $p=10$ & $p=20$ & $p=10$ & $p=20$ & $p=10$ & $p=20$ \\
      \hline
      lme4 &  0.0000 & 0.0000 & 0.0000 & 0.0000 & 0.0000 & 0.0000 & 0.0000 & 0.0000 \\ 
      Meta-lme4 & 0.0000 & 0.0000 & 0.0000 & 0.0000 & 0.0000 & 0.0000 & 0.0000 & 0.0000 \\ 
      IEM &  0.0001 & 0.0071 & 0.0004 & 0.0002 & 0.0000 & 0.0000 & 0.0000 & 0.0000 \\ 
      DEM ($\gamma = 0.3$) & 0.0000 & 0.0000 & 0.0000 & 0.0000 & 0.0000 & 0.0000 & 0.0000 & 0.0000 \\ 
      DEM ($\gamma = 0.5$) & 0.0000 & 0.0000 & 0.0000 & 0.0000 & 0.0000 & 0.0000 & 0.0000 & 0.0000 \\ 
      DEM ($\gamma = 0.7$) & 0.0000 & 0.0000 & 0.0000 & 0.0000 & 0.0000 & 0.0000 & 0.0000 & 0.0000 \\
      \hline
      & \multicolumn{8}{c|}{$K = 20$} \\
      \hline
      & \multicolumn{4}{c|}{$m = 10^4, n = 10^6$} & \multicolumn{4}{c|} {$m = 10^5, n = 10^7$} \\
      \hline
      & \multicolumn{2}{c|}{$q = 3$} & \multicolumn{2}{c|} {$q = 6$} & \multicolumn{2}{c|}{$q = 3$} & \multicolumn{2}{c|} {$q = 6$} \\
      \hline
      & $p=10$ & $p=20$ & $p=10$ & $p=20$ & $p=10$ & $p=20$ & $p=10$ & $p=20$ \\
      \hline
      lme4 & 0.0000 & 0.0000 & 0.0000 & 0.0000 & 0.0000 & 0.0000 & 0.0000 & 0.0000 \\ 
      Meta-lme4 & 0.0001 & 0.0001 & 0.0001 & 0.0001 & 0.0000 & 0.0000 & 0.0000 & 0.0000 \\ 
      IEM &  0.0001 & 0.0002 & 0.0008 & 0.0018 & 0.0001 & 0.0002 & 0.0001 & 0.0001 \\ 
      DEM ($\gamma = 0.3$) &  0.0000 & 0.0000 & 0.0000 & 0.0000 & 0.0000 & 0.0000 & 0.0000 & 0.0000 \\ 
      DEM ($\gamma = 0.5$) &  0.0000 & 0.0000 & 0.0000 & 0.0000 & 0.0000 & 0.0000 & 0.0000 & 0.0000 \\ 
      DEM ($\gamma = 0.7$) &  0.0000 & 0.0000 & 0.0000 & 0.0000 & 0.0000 & 0.0000 & 0.0000 & 0.0000 \\ 
      \hline      
      \hline
      & \multicolumn{8}{c|}{\textbf{Variances of random effects}  (diagonal elements of $\Sigma$)} \\      
      \hline        
      & \multicolumn{8}{c|}{$K = 10$} \\
      \hline
      & \multicolumn{4}{c|}{$m = 10^4, n = 10^6$} & \multicolumn{4}{c|} {$m = 10^5, n = 10^7$} \\
      \hline
      & \multicolumn{2}{c|}{$q = 3$} & \multicolumn{2}{c|} {$q = 6$} & \multicolumn{2}{c|}{$q = 3$} & \multicolumn{2}{c|} {$q = 6$} \\
      \hline
      & $p=10$ & $p=20$ & $p=10$ & $p=20$ & $p=10$ & $p=20$ & $p=10$ & $p=20$ \\
      \hline
      lme4      & 0.0000 & 0.0000 & 0.0000 & 0.0000 & 0.0000 & 0.0000 & 0.0000 & 0.0000 \\ 
      Meta-lme4 & 0.0008 & 0.0008 & 0.0018 & 0.0015 & 0.0001 & 0.0001 & 0.0001 & 0.0002 \\ 
      IEM       & 0.0001 & 0.0017 & 0.0001 & 0.0001 & 0.0000 & 0.0000 & 0.0000 & 0.0000 \\ 
      DEM ($\gamma = 0.3$) & 0.0000 & 0.0000 & 0.0000 & 0.0000 & 0.0000 & 0.0000 & 0.0000 & 0.0000 \\ 
      DEM ($\gamma = 0.5$) & 0.0000 & 0.0000 & 0.0000 & 0.0000 & 0.0000 & 0.0000 & 0.0000 & 0.0000 \\ 
      DEM ($\gamma = 0.7$) & 0.0000 & 0.0000 & 0.0000 & 0.0000 & 0.0000 & 0.0000 & 0.0000 & 0.0000 \\ 
      \hline
      & \multicolumn{8}{c|}{$K = 20$} \\
      \hline
      & \multicolumn{4}{c|}{$m = 10^4, n = 10^6$} & \multicolumn{4}{c|} {$m = 10^5, n = 10^7$} \\
      \hline
      & \multicolumn{2}{c|}{$q = 3$} & \multicolumn{2}{c|} {$q = 6$} & \multicolumn{2}{c|}{$q = 3$} & \multicolumn{2}{c|} {$q = 6$} \\
      \hline
      & $p=10$ & $p=20$ & $p=10$ & $p=20$ & $p=10$ & $p=20$ & $p=10$ & $p=20$ \\
      \hline
      lme4               & 0.0000 & 0.0000 & 0.0000 & 0.0000 & 0.0000 & 0.0000 & 0.0000 & 0.0000 \\ 
      Meta-lme4          & 0.0009 & 0.0009 & 0.0022 & 0.0024 & 0.0001 & 0.0001 & 0.0003 & 0.0003 \\ 
      IEM                & 0.0001 & 0.0001 & 0.0002 & 0.0171 & 0.0000 & 0.0001 & 0.0000 & 0.0000 \\ 
      DEM ($\gamma = 0.3$) & 0.0000 & 0.0000 & 0.0000 & 0.0000 & 0.0000 & 0.0000 & 0.0000 & 0.0000 \\ 
      DEM ($\gamma = 0.5$) & 0.0000 & 0.0000 & 0.0000 & 0.0000 & 0.0000 & 0.0000 & 0.0000 & 0.0000 \\ 
      DEM ($\gamma = 0.7$) & 0.0000 & 0.0000 & 0.0000 & 0.0000 & 0.0000 & 0.0000 & 0.0000 & 0.0000 \\ 
      \hline    
      \hline
      & \multicolumn{8}{c|}{\textbf{Covariances of random effects}  (off-diagonal elements of $\Sigma$)} \\
      \hline    
      & \multicolumn{8}{c|}{$K = 10$} \\
      \hline
      & \multicolumn{4}{c|}{$m = 10^4, n = 10^6$} & \multicolumn{4}{c|} {$m = 10^5, n = 10^7$} \\
      \hline
      & \multicolumn{2}{c|}{$q = 3$} & \multicolumn{2}{c|} {$q = 6$} & \multicolumn{2}{c|}{$q = 3$} & \multicolumn{2}{c|} {$q = 6$} \\
      \hline
      & $p=10$ & $p=20$ & $p=10$ & $p=20$ & $p=10$ & $p=20$ & $p=10$ & $p=20$ \\
      \hline
      lme4 & 0.0000 & 0.0000 & 0.0000 & 0.0000 & 0.0000 & 0.0000 & 0.0000 & 0.0000 \\ 
      Meta-lme4 & 0.0006 & 0.0006 & 0.00010 & 0.0009 &  0.0001 & 0.0001 & 0.0001 & 0.0001 \\ 
      IEM & 0.0000 & 0.0010 & 0.0000 & 0.0000 & 0.0000 & 0.0000 & 0.0000 & 0.0000 \\ 
      DEM ($\gamma = 0.3$) & 0.0000 & 0.0000 & 0.0000 & 0.0000 & 0.0000 & 0.0000 & 0.0000 & 0.0000 \\ 
      DEM ($\gamma = 0.5$) & 0.0000 & 0.0000 & 0.0000 & 0.0000 & 0.0000 & 0.0000 & 0.0000 & 0.0000 \\ 
      DEM ($\gamma = 0.7$) & 0.0000 & 0.0000 & 0.0000 & 0.0000 & 0.0000 & 0.0000 & 0.0000 & 0.0000 \\ 
      \hline
      & \multicolumn{8}{c|}{$K = 20$} \\
      \hline
      & \multicolumn{4}{c|}{$m = 10^4, n = 10^6$} & \multicolumn{4}{c|} {$m = 10^5, n = 10^7$} \\
      \hline
      & \multicolumn{2}{c|}{$q = 3$} & \multicolumn{2}{c|} {$q = 6$} & \multicolumn{2}{c|}{$q = 3$} & \multicolumn{2}{c|} {$q = 6$} \\
      \hline
      & $p=10$ & $p=20$ & $p=10$ & $p=20$ & $p=10$ & $p=20$ & $p=10$ & $p=20$ \\
      \hline
      lme4                 & 0.0000 & 0.0000 & 0.0000 & 0.0000 & 0.0000 & 0.0000 & 0.0000 & 0.0000 \\
      Meta-lme4            & 0.0006 & 0.0007 & 0.0015 & 0.0014 & 0.0001 & 0.0001 & 0.0002 & 0.0002 \\
      IEM                  & 0.0000 & 0.0000 & 0.0001 & 0.0040 & 0.0000 & 0.0000 & 0.0000 & 0.0000 \\ 
      DEM ($\gamma = 0.3$) & 0.0000 & 0.0000 & 0.0000 & 0.0000 & 0.0000 & 0.0000 & 0.0000 & 0.0000 \\ 
      DEM ($\gamma = 0.5$) & 0.0000 & 0.0000 & 0.0000 & 0.0000 & 0.0000 & 0.0000 & 0.0000 & 0.0000 \\  
      DEM ($\gamma = 0.7$) & 0.0000 & 0.0000 & 0.0000 & 0.0000 & 0.0000 & 0.0000 & 0.0000 & 0.0000 \\
      \hline
    \end{tabular}
  }%
\end{table}

\begin{table}[ht]
  \caption{The ratio of IEM or DEM and ECME$_0$ log likelihoods averaged over simulation replications. Monte Carlo errors are in parenthesis}  
  \label{tab:loglik}  
  \centering
{\scriptsize
  \begin{tabular}{|r|c|c|c|c|c|c|c|c|}
    \hline
    & \multicolumn{8}{c|}{\textbf{IEM or DEM log likelihood / ECME$_0$ log likelihood }} \\
    \hline
     & \multicolumn{8}{c|}{$K = 10$} \\
    \hline
     & \multicolumn{4}{c|}{$m = 10^4, n = 10^6$} & \multicolumn{4}{c|} {$m = 10^5, n = 10^7$} \\
    \hline
     & \multicolumn{2}{c|}{$q = 3$} & \multicolumn{2}{c|} {$q = 6$} & \multicolumn{2}{c|}{$q = 3$} & \multicolumn{2}{c|} {$q = 6$} \\
    \hline
     & $p=10$ & $p=20$ & $p=10$ & $p=20$ & $p=10$ & $p=20$ & $p=10$ & $p=20$ \\
    \hline
    IEM                &  1.00 (0.00) & 1.00 (0.00) & 1.00 (0.00) & 1.00 (0.00) & 1.00 (0.00) & 1.00 (0.00) & 1.00 (0.00) & 1.00 (0.00) \\ 
  DEM ($\gamma = 0.3$) & 1.00 (0.00) & 1.00 (0.00) & 1.00 (0.00) & 1.00 (0.00) & 1.00 (0.00) & 1.00 (0.00) & 1.00 (0.00) & 1.00 (0.00)\\ 
  DEM ($\gamma = 0.5$) & 1.00 (0.00) & 1.00 (0.00) & 1.00 (0.00) & 1.00 (0.00) & 1.00 (0.00) & 1.00 (0.00) & 1.00 (0.00) & 1.00 (0.00)\\ 
  DEM ($\gamma = 0.7$) & 1.00 (0.00) & 1.00 (0.00) & 1.00 (0.00) & 1.00 (0.00) & 1.00 (0.00) & 1.00 (0.00) & 1.00 (0.00) & 1.00 (0.00) \\
    \hline
    & \multicolumn{8}{c|}{$K = 20$} \\
    \hline
    & \multicolumn{4}{c|}{$m = 10^4, n = 10^6$} & \multicolumn{4}{c|} {$m = 10^5, n = 10^7$} \\
    \hline
    & \multicolumn{2}{c|}{$q = 3$} & \multicolumn{2}{c|} {$q = 6$} & \multicolumn{2}{c|}{$q = 3$} & \multicolumn{2}{c|} {$q = 6$} \\
    \hline
    & $p=10$ & $p=20$ & $p=10$ & $p=20$ & $p=10$ & $p=20$ & $p=10$ & $p=20$ \\
    \hline
    IEM                & 1.00 (0.00) & 1.00 (0.00) & 1.00 (0.00) & 1.00 (0.00) & 1.00 (0.00) & 1.00 (0.00) & 1.00 (0.00) & 1.00 (0.00)\\ 
  DEM ($\gamma = 0.3$) & 1.00 (0.00) & 1.00 (0.00) & 1.00 (0.00) & 1.00 (0.00) & 1.00 (0.00) & 1.00 (0.00) & 1.00 (0.00) & 1.00 (0.00) \\ 
  DEM ($\gamma = 0.5$) & 1.00 (0.00) & 1.00 (0.00) & 1.00 (0.00) & 1.00 (0.00) & 1.00 (0.00) & 1.00 (0.00) & 1.00 (0.00) & 1.00 (0.00) \\ 
  DEM ($\gamma = 0.7$) & 1.00 (0.00) & 1.00 (0.00) & 1.00 (0.00) & 1.00 (0.00) & 1.00 (0.00) & 1.00 (0.00) & 1.00 (0.00) & 1.00 (0.00)\\ 
    \hline
  \end{tabular}
}
\end{table}

\begin{table}[ht]
  \caption{Root mean square error \eqref{eq:met2} in estimation of  fixed effects ($\betab$), variances of random effects (diagonal elements of $\Sigma$), error variance ($\tau^2$), and covariances of random effects (off-diagonal elements of $\Sigma$) averaged over all replications. The maximum Monte Carlo errors are of the order $10^{-3}$, $10^{-4}$, $10^{-3}$, and $10^{-4}$, respectively. The subscripts $1, \ldots, 6$ represent \emph{Action}, \emph{Children} $-$ \emph{Action}, \emph{Comedy} $-$ \emph{Action}, \emph{Drama} $-$ \emph{Action}, \emph{popularity}, and \emph{previous} predictors} 
  \label{tab:rcovs}
  \centering
{\scriptsize
  \begin{tabular}{|r|cccccccc|}
  \hline
  & $\beta_{\text{Action}}$ & $\beta_{\text{Children $-$ Action}}$ & $\beta_{\text{Comedy $-$ Action}}$ & $\beta_{\text{Drama $-$ Action}}$ & $\beta_{\text{popularity}}$ & $\beta_{\text{previous}}$ & & \\
  \hline
  lme4                 &  0.0000 & 0.0000 & 0.0000 &  0.0000 &  0.0000 &  0.0000   & & \\ 
  Meta-lme4            &  0.0000 & 0.0002 & 0.0001 &  0.0001 &  0.0000 &  0.0000   & & \\ 
  IEM                  &  0.0000 & 0.0000 & 0.0000 &  0.0000 &  0.0000 &  0.0000   & & \\ 
  DEM ($\gamma = 0.3$) &  0.0000 & 0.0000 & 0.0000 &  0.0000 &  0.0000 &  0.0000   & & \\ 
  DEM ($\gamma = 0.5$) &  0.0000 & 0.0000 & 0.0000 &  0.0000 &  0.0000 &  0.0000   & & \\ 
  DEM ($\gamma = 0.7$) &  0.0000 & 0.0000 & 0.0000 &  0.0000 &  0.0000 &  0.0000   & & \\ 
  \hline
  \hline
  & $\sigma^2_{\text{Action}}$ & $\sigma^2_{\text{Children $-$ Action}}$ & $\sigma^2_{\text{Comedy $-$ Action}}$ & $\sigma^2_{\text{Drama $-$ Action}}$ & $\sigma^2_{\text{popularity}}$ & $\sigma^2_{\text{previous}}$ & $\tau^2$ & \\
  \hline
  lme4 &  0.0000 &  0.0000 &  0.0000 &  0.0000 &  0.0000 &  0.0000 &  0.0006 & \\ 
  Meta-lme4 &  0.0001 &  0.0002 &  0.0001 &  0.0000 &  0.0000 &  0.0001 &  0.0007 & \\ 
  IEM &   0.0000 &  0.0000 &  0.0000 &  0.0000 &  0.0000 &  0.0000 &  0.0010 & \\ 
  DEM ($\gamma = 0.3$) & 0.0000 &  0.0000 &  0.0000 &  0.0000 &  0.0000 &  0.0000 &  0.0008 & \\ 
  DEM ($\gamma = 0.5$) & 0.0000 &  0.0000 &  0.0000 &  0.0000 &  0.0000 &  0.0000 &  0.0007 & \\ 
  DEM ($\gamma = 0.7$) & 0.0000 &  0.0000 &  0.0000 &  0.0000 &  0.0000 &  0.0000 &  0.0006 & \\ 
   \hline
    \hline
    &  $\sigma_{12}$ & $\sigma_{13}$ & $\sigma_{14}$ & $\sigma_{15}$ & $\sigma_{16}$ & $\sigma_{23}$ & $\sigma_{24}$ & $\sigma_{25}$ \\ 
    \hline
    lme4      &  0.0000 &  0.0000 &  0.0000 &  0.0000 &  0.0000 &  0.0000 &  0.0000 &  0.0000 \\ 
    Meta-lme4 &  0.0001 &  0.0001 &  0.0000 &  0.0000 &  0.0001 &  0.0001 &  0.0001 &  0.0000  \\ 
    IEM       &   0.0000 &  0.0000 &  0.0000 &  0.0000 &  0.0000 &  0.0000 &  0.0000 &  0.0000 \\ 
    DEM ($\gamma = 0.3$) &  0.0000 &  0.0000 &  0.0000 &  0.0000 &  0.0000 &  0.0000 &  0.0000 &  0.0000\\ 
    DEM ($\gamma = 0.5$) &  0.0000 &  0.0000 &  0.0000 &  0.0000 &  0.0000 &  0.0000 &  0.0000 &  0.0000\\ 
    DEM ($\gamma = 0.7$) &  0.0000 &  0.0000 &  0.0000 &  0.0000 &  0.0000 &  0.0000 &  0.0000 &  0.0000\\ 
    \hline
    &  $\sigma_{26}$ & $\sigma_{34}$ & $\sigma_{35}$ & $\sigma_{36}$ & $\sigma_{45}$ & $\sigma_{46}$ & $\sigma_{56}$ & \\ 
    \hline
    lme4      &  0.0000 &  0.0000 &  0.0000 &  0.0000 &  0.0000 &  0.0000 &  0.0000 &   \\ 
    Meta-lme4 &  0.0001 &  0.0000 &  0.0000 &  0.0000 &  0.0001 &  0.0001 &  0.0000 &    \\ 
    IEM       &   0.0000 &  0.0000 &  0.0000 &  0.0000 &  0.0000 &  0.0000 &  0.0000 &  \\ 
    DEM ($\gamma = 0.3$) &  0.0000 &  0.0000 &  0.0000 &  0.0000 &  0.0000 &  0.0000 &  0.0000 & \\ 
    DEM ($\gamma = 0.5$) &  0.0000 &  0.0000 &  0.0000 &  0.0000 &  0.0000 &  0.0000 &  0.0000 & \\ 
    DEM ($\gamma = 0.7$) &  0.0000 &  0.0000 &  0.0000 &  0.0000 &  0.0000 &  0.0000 &  0.0000 & \\ 
    \hline
  \end{tabular}
}%
\end{table}

\begin{table}[ht]
  \caption{The ratio of IEM or DEM log likehood over ECME$_0$ log likehood averaged over all replications. Monte Carlo errors are in parenthesis}
  \label{tab:rllk}
  \centering
  {\scriptsize
\begin{tabular}{|cccc|}
  \hline
   IEM  &   DEM ($\gamma = 0.30$)  &  DEM ($\gamma = 0.50$)  &  DEM ($\gamma = 0.70$)  \\ 
  \hline
  1.0000 (0.0000) & 1.0000 (0.0000) & 1.0000 (0.0000) & 1.0000 (0.0000) \\ 
   \hline
\end{tabular}    
}%
\end{table}

\end{document}